\documentclass[pra,groupedaddress,superscriptaddress,amsmath,amssymb,a4paper,twocolumn,floatfix]{revtex4}

\usepackage{geometry}
\usepackage{graphicx}
\usepackage{verbatim}
\usepackage{float}
\usepackage{graphics}
\usepackage{mathrsfs}
\usepackage{amssymb}
\usepackage{amsmath}
\usepackage{amsthm}
\usepackage{bm}
\usepackage{hyperref}
\usepackage{braket}
\usepackage{mathtools}
\usepackage{multirow}
\usepackage{pst-node}
\usepackage{amsbsy}
\usepackage{tikz}
\usepackage{float}
\usepackage{natbib}
\usepackage[english]{babel}

\usetikzlibrary{calc}
\usepackage[nodisplayskipstretch]{setspace}

\def\VR{\kern-\arraycolsep\strut\vrule &\kern-\arraycolsep}

\addcontentsline{toc}{chapter}{References}

\begin{document}

\title{Classification of four-qubit entangled states via  Machine Learning}

\author{S. V. Vintskevich}
\affiliation{Technology Innovation Institute, Abu Dhabi, United Arab Emirates}

\author{N. Bao}
\affiliation{Brookhaven National Laboratory, Upton NY 11973, USA}

\author{A. Nomerotski}
\affiliation{Brookhaven National Laboratory, Upton NY 11973, USA}

\author{P. Stankus}
\affiliation{Brookhaven National Laboratory, Upton NY 11973, USA}

\author{D.A. Grigoriev}
\affiliation{LLP Eqvium, Almaty, Kazakhstan}
\begin{abstract}
We apply the support vector machine (SVM) algorithm to derive a set of entanglement witnesses (EW) to identify entanglement patterns in families of four-qubit states.
The effectiveness of SVM for practical EW implementations stems from the coarse-grained description of families of equivalent entangled quantum states.
The equivalence criteria in our work is based on the stochastic local operations and classical communication (SLOCC) classification and the description of the four-qubit entangled Werner states.
We numerically verify that the SVM approach provides an effective tool to address the entanglement witness problem when the coarse-grained description of a given family state is available.
We also discuss and demonstrate the efficiency of nonlinear kernel SVM methods as applied to four-qubit entangled state classification.
\end{abstract}


\maketitle

\section{Introduction}
The ability to quantify, detect and analyze the structure of quantum entanglement \cite{Horodecki2009} is essential for quantum computation \cite{Jozsa2003,Cluster2001,Zhang2006}, quantum communication \cite{Wei2022, Hu2021,Lloyd2003}, quantum networks \cite{McCutcheon2016,Illiano2022,Jessica2022} and quantum metrology \cite{Toth2012,Hyllus2012}.
Moreover, an improper ''amount" of entanglement, incorrect structure or pattern of multipartite entangled state, or action of quantum noise could all severely affect the overall efficiency of given quantum computation tasks\cite{Gross2009} or performance of given quantum protocols, such as that of entanglement purification \cite{Dr2007}.
Thus, it is crucial to detect and describe the structure of entanglement in a set of states to maximize the efficiency of a given protocol.

As a specific case of this, one can attempt to detect the presence of entanglement in particular families of states relevant in specific quantum protocols; note that the success probability of the protocol may vary when one uses different families of entangled states \cite{Eisert2020,Kliesch2021,Guhne2021}. 
A characteristic example of such a difference is that between the multipartite GHZ- and W- type states. 
The GHZ-type states are fragile against losses and are more utilized in quantum information sharing protocols, whereas W-type states are robust against the noise and are used in multi-party quantum network protocols \cite{Cruz2019}.
For instance, in the case of tripartite W-state it is easy to verify that each pair of qubits in this state is in an entangled state in contrast to the tripartite GHZ-type state.
This particular example emphasizes  robustness of the W-state, e.g. we can assume that the third qubit can be traced out to emulate losses.

The entanglement witness (EW) technique is one of the most common, effective and practical methods to detect the presence of entanglement for a given {\it{multipartite}} quantum state, see, e.g.\cite{Ghne2009, Horodecki2009}, and references therein.
In the present work, we analyze this technique applied to arbitrary four-qubit systems.
The essence of the EW technique is briefly summarized below.
By finding a specific Hermitian operator, called the entanglement witness $\hat{W}$,
one calculates a linear functional (EW functional), which maps a given multipartite quantum state described by density operator $\hat{\varrho}$ into a real number \footnote{The usage of density operators allows the accommodation of both mixed and pure states.} with a property that the EW functional has a non-negative value for all separable states $\hat{\varrho}_{sep.}$ and that there exists a particular set of entangled states for which the EW functional has negative values \cite{HORODECKI19961,TERHAL2000319,Horodecki2009}. 

More specifically, let us formally denote a set of arbitrary multi-qubit states as: $S_{n_{q}}: = \{\hat{\varrho}_{n_{q}} \in \mathcal{T}(\mathcal{H}_{n_{q}})| \ {\rm{tr}}\left(\hat{\varrho}_{n_{q}}\right) = 1, \hat{\varrho}_{n_{q}}\geq O\}$. 
We denoted $\mathcal{T}\left(\mathcal{H}_{n_{q}}\right)$ as a linear space of trace class operators acting on Hilbert space $\mathcal{H}_{{n}_{q}}$. The condition $\hat{\varrho}_{n_{q}}\geq O$ states that operator $\hat{\varrho}$ is a positive operator \cite{Heinosaari2009}.
Mathematically, the EW functional will have the following property:
\begin{equation}\label{witness_general_def}
    {\rm{tr}}\left(\hat{\varrho}_{sep.} \hat{W}\right)\geq 0, \forall \hat{\varrho}_{sep.} \in S; \ {\rm{tr}}\left(\hat{\varrho}_{ent.} \hat{W}\right) < 0,
\end{equation}
where we denote $S$ as a subset of all separable states. By definition \cite{Horodecki2009,Heinosaari2009} a state $\hat{\varrho}_{sep.}$ is separable if and only if it can be represented as a convex combination of factorized states: $\hat{\varrho} = \sum_{i}^{k}p_{i}\hat{\varrho}^{i}_{1}\otimes\dots\otimes\hat{\varrho}^{i}_{N}$; where each $\hat{\varrho}^{i}$ is a state of the $i$-th qubit subsystem, a density operator acting on a subspace $\mathcal{H}_{i}$ of a N-partite quantum system with joint space $\mathcal{H} = \mathcal{H}_{i}\otimes \dots  \mathcal{H}_{N}$; $\sum_{i=1}^{k} p_{i} = 1, \ k \leq {\rm{dim}}\mathcal{H}^{2}$.
Thus, the set of all separable states is a convex subset of all states $S_{{n_{q}}}$ defined above with respect to the trace norm \cite{Heinosaari2009}, while, by definition, the entangled states are those states that are not separable.

Note that the entanglement witness operator $\hat{W}$, which can detect the entanglement for a given state $\hat{\varrho}_{ent}$ is not universal. 
There are always {\it{other}} entangled states $\hat{\sigma}_{ent}$ such that ${\rm{tr}}(\hat{W}\hat{\sigma}_{ent})>0$, but at the same time ${\rm{tr}}(\hat{W}\hat{\varrho}_{ent})<0$. Unfortunately, one cannot determine the entanglement witness operator for all possible entangled states for an arbitrary multipartite quantum system \cite{Horodecki2009,Heinosaari2009}.
It was shown that the problem of a general description of all entangled states, pure and mixed, for multipartite quantum systems does not have a solution, see e.g.  \cite{Ioannou2004,Ioannou2006,Ioannou2007,bengtsson2017geometry}.
However, in some cases, the aforementioned disadvantage might be partially overcome.
The general idea is to split the quantum states into specific families - sets of states that may share certain symmetries or specific structure.
Another possibility is if the quantum states can be created via a specific protocol that maps a certain set of states to another set that can be mathematically described via a specific parametrization.

The idea of splitting sets of entangled quantum states (generally, an infinite set of states) into families with some inner mathematical structure can be termed  {\it{a coarse-grained classification}}.
For instance one can specify equivalence classes - families of the pure entangled states. Each state within a given family can be transformed into another state of the same family with nonzero probability through local operations and classical communications (LOCC) \cite{Dur2000,Verstraete2002}. 
If one can describe or specify a structure of entangled states one may expect that the complexity of finding entanglement witnesses for such set of states will be significantly reduced.
The mentioned above GHZ and W states are members of two distinguishable families of states that cannot be converted into one another by any (SLOCC) as was emphasized in \cite{Verstraete2002}. 
Other possible approaches include inductive entanglement classification \cite{Lamata2007}, entanglement classification with matrix product states \cite{Sanz2016} and coarse graining of entanglement classes in $2 \times m \times n$ systems \cite{Gachechiladze2018}.

Here we aim to employ the coarse-grained classifications of four-qubit quantum states, following the results of \cite{Verstraete2002}, to construct a set of entanglement witness operators. 
We avoid difficulties of analytical derivation of the entanglement witness operators by approaching the problem numerically. 
The core of our numerical analysis is based on the well-known Support Vector Machine (SVM) method in machine-learning (ML) \cite{cortes1995support,bishop2007}.
The SVM-based algorithm is designed to detect the presence of entanglement not only in arbitrary four-qubit states but also to assign this state to a particular family of states. 
To train our SVM model we sample a data set of quantum states for each class of entangled and separable states based on this coarse-grained classification. We are sampling $20000$ states for each class of entangled states, including separable states.

It is worth emphasizing that in recent years, machine learning-based methods have demonstrated remarkable efficiency in application to various areas of quantum physics \cite{Carleo2017,Torlai2018,RMP2019ML,LVGF}.
For instance, others have used the neural networks \cite{Ma2018,Lu2018} and also the SVM \cite{zhu2021} to find EW operators, which efficiently distinguish between separable and entangled states of a particular type. 

The simplest linear SVM approach can classify quantum states if they belong to one of two classes (entangled and probably separable) by computing the {\it{decision function}} and constructing a {\it{decision boundary}}. 
Thus, the SVM approach directly corresponds to the EW problem. Note that SVM fits our problem naturally, as quantum states can be equivalently considered as vectors in $\mathcal{T}\left(\mathcal{H}_{n_{q}}\right)$ space.
Consequently, the SVM approach allows one to find a hyper-plane in the $\mathcal{T}\left(\mathcal{H}_{n_{q}}\right)$, which corresponds to the entanglement witness.
We describe our approach and results of its application in Section \ref{SEC2}.
Appendix \ref{appendix-a} provides a detailed description of the SLOCC classification of four-qubit states used in this work.

\section{Entanglement witnesses for four-qubit states and application of SVM}\label{SEC2}

In this section we construct a set of EW operators by considering a set of four qubit states ($n_{q} = 4$).
As the first step, let us provide a general mathematical description for the linear space of trace class operators $\mathcal{T}\left(\mathcal{H}_{n_{q}}\right)$ in the case of an arbitrary number of qubits $n_{q}$.
The space $\mathcal{T}\left(\mathcal{H}_{n_{q}}\right)$ is endowed with Hilbert-Schmidt inner product for any two operators $\hat{R}_{1},\hat{R}_{2}$: $\langle \hat{R}_{1},\hat{R}_{2} \rangle_{{\rm{HS}}} = {\rm{tr}}\left(\hat{R}_{1}^{\dagger} \hat{R}_{2}\right)$, which induces the norm $||R||_{HS} = \sqrt{\langle R^{\dagger}R\rangle_{HS}}$. Let us choose the standard multi-qubit Pauli basis $\{\hat{\mathbf{I}}/\sqrt{2},\hat{\sigma}_{x}/\sqrt{2},\hat{\sigma}_{y}/\sqrt{2},\hat{\sigma}_{z}/\sqrt{2}\}^{\otimes {n_{q}}}$ as a self-adjoint orthonormal basis for $\mathcal{T}\left(\mathcal{H}_{n_{q}}\right)$. 
For simplicity we denoted a particular basis operator as $\hat{E}_{i}, i = 0, 4^{{n_{q}}}-1$ assuming $ \hat{E}_{0} = \hat{\mathbf{I}}^{\otimes n_{q}}$, ${\rm{tr}}\left(E_{i}\right) = 0, i\neq0$ and $\langle \hat{E}_{i}\hat{E}_{j} \rangle_{HS} = \delta_{ij}$.
Thus, an arbitrary operator $\hat{R}$ can be represented as a vector $\vec{r}$:
\begin{eqnarray}\label{vec_representation}
\hat{R} = \sum_{j}r_{j}\hat{E}_{j},r_{k} = {\rm{tr}}\left(\hat{E}_{k}\hat{R}\right), \vec{r} = (r_{0}, \dots r_{4^{n_{q}} - 1})
\end{eqnarray}
If operator $\hat{R}$ is Hermitian all elements of a corresponding vector $\vec{r}$ are real numbers and $||\hat{R}||_{HS} = ||\vec{r}||_{e}$, the norm $||\cdot||_{e}$ is the standard Euclidean norm of a vector.
Consequently, the Hilbert-Schmidt inner product of two Hermitian operators
$\langle R_{1},R_{2} \rangle_{{\rm{HS}}}$ corresponds to the standard "Euclidean" inner product $(\vec{r}_{1},\vec{r}_{2})$.
The vector representation emphasizes the direct correspondence between EW problems and the linear SVM method, and, indeed, the EW problems can be viewed as problems of classification.
We consider two classes: the set of all separable states and a subset of entangled states.
In the vector representation an EW operator and an arbitrary state can be denoted $\hat{W} \leftrightarrow \vec{w}$ and $\hat{\varrho} \leftrightarrow \vec{\varrho}$ respectively, in accordance with \eqref{vec_representation}. 
Thus, the decision function used in SVM can be written as follows:
\begin{eqnarray}\label{svm_simpliest_decision_function}
t_{\hat{W}}\left(\hat{\varrho}\right) = \begin{cases}
-1, & {\text{if}} \  (\vec{w},\vec{\rho}) < 0 \\
1,  & {\text{if}} \  (\vec{w},\vec{\rho}) > 0,
\end{cases}
\end{eqnarray}
where we explicitly write the linear form of EW functional $f_{\hat{W}}(\varrho) = (\vec{w},\vec{\rho}) \equiv {\rm{tr}}(\hat{W}\hat{\varrho})$.
Accordingly, a training data set $DS = \{\hat{\varrho}_{j}\}_{j = 1}^{N_{data}}$ consists of separable and entangled states in vector representation labeled with $t_{\hat{W}}\left(\hat{\varrho}_{sep.}\right) = 1 $ and $t_{\hat{W}}\left(\hat{\varrho}_{ent.}\right) = -1 $ respectively, whereas a decision hyperplane is defined by $f_{\hat{W}}(\hat{\varrho}) = 0$. 

\begin{figure}[ht!]
\begin{center}
\includegraphics[width=1\linewidth]{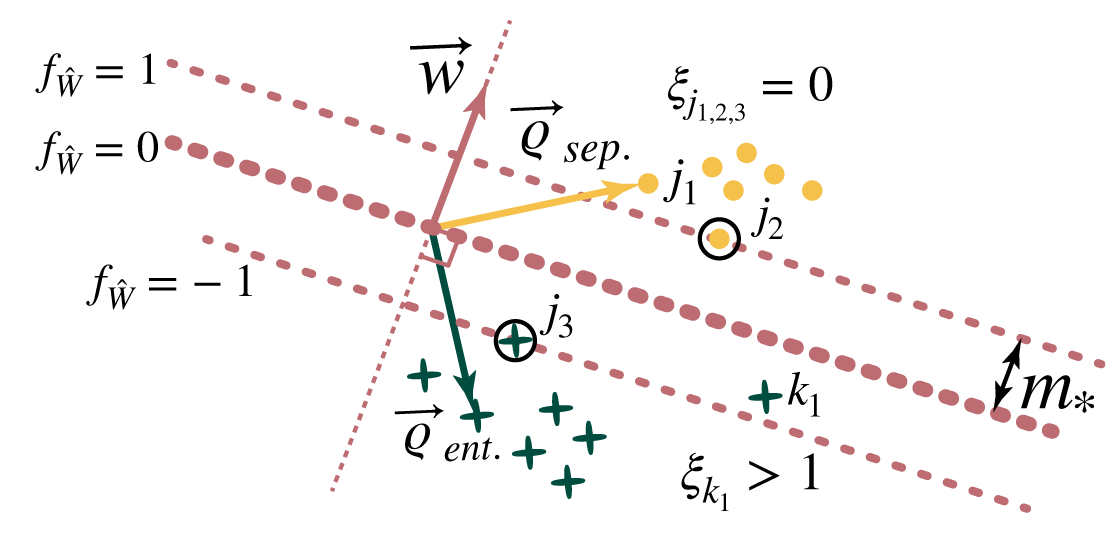}
\caption{Illustration of SVM approach in two dimensions. 
The distance from a given sample state $\hat{\varrho}_{j}\leftrightarrow\vec{\varrho}_{j}$ to a decision boundary $f_{\hat{W}} = 0$ is given by $m_{j} = t_{j}f_{j}/||\vec{w}||$.
Mathematically, the value of an optimal margin value $m_{*}$ can be found by minimizing the hinge error function \eqref{hinge_loss}.
For simplicity we denote $\xi_{j} = {\text{max}
}(0, 1 - f_{j}t_{j})$.
The location of the decision boundary is determined by a subset of the data points, known as {\it{support vectors}}, which is located on the margin boundary ($f_{\hat{W}} = 1$ and $f_{\hat{W}} = -1$, respectively).
The support vectors are marked by the circles. Thus, for the states that are on the correct side of the decision boundary $\xi_{j} = 0$ and for the outliers $\xi_{j} = 1 - t_{j}f_{j}$.
Note that adding new samples that lie outside of the optimal margin region will not affect the decision boundary.}
\label{SVM_GEOMETRY}
\end{center}
\end{figure}

For a given sampled state of a train data set $\hat{\varrho}_{j}$ we will simplify the notation: $f_{\hat{W}}(\hat{\varrho}_{j}) \equiv f_{j} $ and $t_{\hat{W}}(\hat{\varrho}_{j}) \equiv t_{j} $.
The SVM training objective is to find a decision boundary by maximizing the margin $m$, which is the smallest distance between the decision hyperplane (i.e. the boundary) and the closest quantum states, also named {\it{support vectors}}, from the training data set. 
We illustrate the SVM approach in Figure (\ref{SVM_GEOMETRY}). 

To find the optimal margin $m_{*}$ one can consider {\it{minimization}} of the following objective function $\underset{\vec{w}, {\text{DS}}}{\text{argmin}} L(\vec{w},DS)$ named the {\it{hinge}} error function.
The reader may find a comprehensive description in \cite{HandsOn, bishop2007}:
\begin{eqnarray}\label{hinge_loss}
L(\vec{w},DS) =   \sum_{j}^{N_{data}}{\text{max}}\left(0, 1- f_{j}t_{j}\right) + \lambda ||\vec{w}||^{2}_{e}.
\end{eqnarray}
The parameter $\lambda$ plays the role of regularization parameter to control the model's accuracy and generalization abilities.

Due to the given vector representation (\ref{vec_representation}), the linear structure of entanglement witness functional (\ref{witness_general_def}) and relatively small dimensionality of the linear space of operators  $\mathcal{T}(\mathcal{H}_{{n_{q}}})$ (in our case $n_{q} = 4$, $n_{f} ={\rm{dim}} (\mathcal{T}(\mathcal{H}_{{n_{q}}})) = 256$), the SVM approach is computationally feasible. We denoted the dimension of a feature space as $n_{f}$, which in our case coincides with the dimension of space  $\mathcal{T}(\mathcal{H}_{{n_{q}}})$.
For instance, a rough estimate of the SVM computational complexity is $O(N_{data} n_{f})$, based on the implementation of \textsc{SciKit Learn} python package \cite{HandsOn}.
Thus, the SVM computational complexity can be estimated by $O(N_{data} 4^{{n}_{q}})$.
Therefore, the SVM is a good match for analyzes of complex but relatively small data sets.

In addition, the case of four qubits has another advantage, the SLOCC classification of entangled states.
This coarse-grained classification of four-qubit entangled states further reduces the complexity of the EW problem and allows efficient application of SVM.
In the next subsection \ref{Verstratecore} we focus on this classification following the results of \cite{Verstraete2002} and applying it to the construction of EW operators.
The subsection \ref{SVM_core} provides results of the SVM approach for construction of the EW operators and entanglement detection of arbitrary four-qubit states.

\subsection{Classification of four-qubit states}
\label{Verstratecore}

It was shown in \cite{Verstraete2002} how one can classify all pure states of four qubits into distinguishable classes of entangled {\it{pure}} states.
The authors specified nine equivalent classes of this form.
The equivalence criteria were determined with respect to the stochastic local quantum operations assisted by classical communication (SLOCC) applied to a particular quantum state.
More specifically, states in a set can be treated as equivalent if any state for this set can be transformed into {\it{any other state}} with non-zero probability by means of LOCC.
Note that in the present analysis we restrict ourselves to only local operations that are unitary transformations.
A local unitary transformation of arbitrary four-qubit state can be written as:
\begin{equation}\label{SLOCC_general}
    \hat{\varrho}_{'} = \hat{U}_{1}\otimes\hat{U}_{2}\otimes\hat{U}_{3}\otimes\hat{U}_{4}~\hat{\varrho}~\hat{U}_{1}^{\dagger}\otimes\hat{U}_{2}^{\dagger}\otimes\hat{U}_{3}^{\dagger}\otimes\hat{U}_{4}^{\dagger},
\end{equation}
where $\hat{U}$ are arbitrary unitary operators acting on Hilbert space of a single qubit $\mathcal{H}_{2}$.
Operationally, the states belonging to the same class can be used in a given quantum protocol but they will have a different protocol efficiency.
We list the explicit state classification in Appendix \ref{appendix-a} based on results of \cite{Verstraete2002}. 
We denote these {\it{pure states}} via ket - vectors $\ket{G},\ket{E}_{i}, i = 1,\dots,8, \ket{F}_{0}$, where $\ket{G}$ represents a generic state, as $\ket{E}_{i}$ we denoted "specific" classes of entangled states and $\ket{F}_{0}$ is a factorized state, so is a class member of separable four-qubit states. 

Some of the states, like the class representatives in \eqref{G} - \eqref{E8}, are parametrized with complex numbers, e.g. the generic states such as $\ket{G}$ in \eqref{G} and states $\ket{E}_{i}, i = 1,5$ in  \eqref{E1}-\eqref{E5}.
However, some classes are represented just by a single pure state.
To form a data set needed to train our SVM model we sample the complex numbers for each class $\{{E}_{i}, i = 1,5\}$.
In addition, we sample the random unitary matrices $\hat{U}$ to increase the diversity of a particular data set, in accordance with \eqref{SLOCC_general}. 

The sampling algorithm of random unitary matrices is implemented via \textsc{QuTip python package} \cite{QUTIP2012,QUTIP2013} with a slight modification as explained below.
We took the standard {\textsc{QuTip}} function {\texttt{rand}}$\_${\texttt{unitary}} and added an additional parameter $\varepsilon$ to control the value of $U_{\varepsilon}$.
For this we assumed that:
\begin{eqnarray}\label{U_sampling}
U_{\varepsilon} = \rm{exp}
\left(i\varepsilon H\right),
\end{eqnarray}
where $H$ is a random $2\times2$ full rank Hermitian matrix.
The sampling of $H$ is done with another {\textsc{QuTip}} function, {\texttt{rand}}$\_${\texttt{herm}}.

\begin{figure}[tb]
\begin{center}
\includegraphics[width=0.72\linewidth]{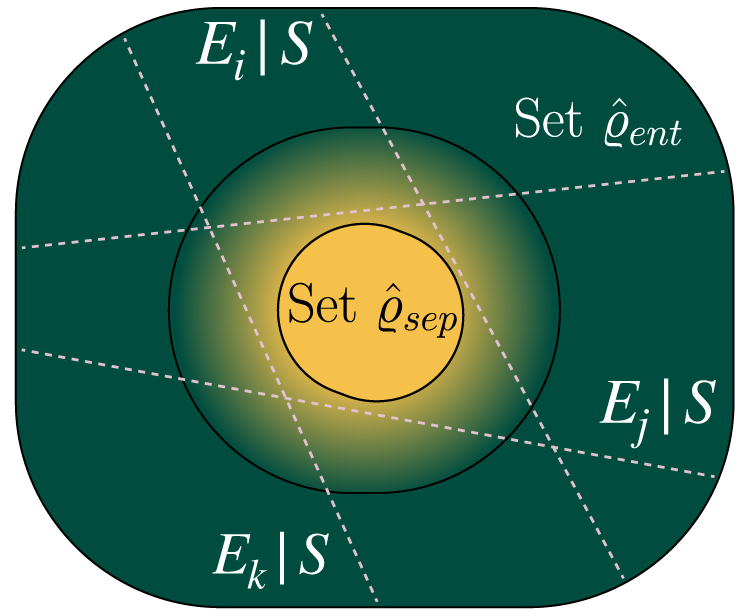}
\caption{Schematic representation of results obtained with the SVM approach.
The decision boundaries between the set of separable states $S$ and each class of entangled states ${E}_{i}$ are represented by dashed lines.
Each line corresponds to a particular class and is specified by a vector representation $\vec{w}_{i}$ of the entanglement witness operator $\hat{W}_{i}$ in accordance with Eqs. \eqref{witness_general_def} and Eqs. \eqref{svm_simpliest_decision_function}}.
\label{GEOMETRY}
\end{center}
\end{figure}

Summarizing, for each SLOCC class of four-qubit states we create $N_{data} = 20000$ samples of pure states, including factorized states, utilizing the LOCC  transformations for each state \eqref{SLOCC_general} with random unitary operators in accordance with Eq. \eqref{U_sampling}.
We normalize each sample by replacing $\hat{\varrho}_{E_{i}} \rightarrow \hat{\varrho}_{E_{i}}/{\rm{tr}}\left(\hat{\varrho}_{E_{i}}\right)$.
As the next step, we find the EW operators via a SVM algorithm, which can distinguish separable four-qubit states $\hat{\varrho}_{sep}$ and each class of entangled states represented in families  ${E}_{i}, i = 1,\dots 8$.
Figure \ref{GEOMETRY} illustrates the expected result.

The set of separable states $S$ is a closed convex set and any separable state can be represented as a convex hull of factorized states.
We directly employ this property to create a data set of separable states to train and validate our results achieving high accuracy, see subsection \ref{SVM_core}.
To create a single separable state we sample $4 * 4^{4}$ random factorized unitary operators (matrices): $\hat{U}_{f}^{(4)} \equiv \hat{U}_{1} \otimes \hat{U}_{2} \otimes \hat{U}_{3}\otimes \hat{U}_{4}$.
Each $2\rm{x}2$ single qubit unitary operator $\hat{U}$ is distributed according to the Haar measure \cite{FasiHaar}.
As a result one can get a random pure factorized state $\ket{F}' = \hat{U}_{f}^{(4)} \ket{F}_{0}$.
The need for $4 * 4^{4}$ unitary operators is driven by the requirement of Carathéodory's theorem for the convex sets and dimension of state space $\mathcal{H}_{n_{q}}$, see \cite{Heinosaari2009} for a detailed discussion.
To sample an arbitrary separable four-qubit state we construct a convex hull: $\hat{\varrho}_{S} = \sum_{i = 1}^{4^{4}}p_{i}\ket{F}'_{i}\bra{F}'_{i}$, where the probability distribution $\{p_{i}\}$ is generated from the Dirichlet distribution \cite{bishop2007}.
Note that it is possible to train SVM and find the EW operator by using only pure factorized states.
It follows directly from the definition of separable states and properties of entanglement witnesses.
We will return to this topic in the next subsection.

Summarizing, our central goal is to demonstrate that the SVM approach allows one to classify and detect the entanglement for an {\it{arbitrary}} four-qubit state, not only for pure states.
In the next subsection we confirm that the presented above SVM approach can achieve this goal with high accuracy.

\subsection{Constructing  Entanglement Witness operators via Linear Support Vector Machines}\label{SVM_core}

To implement the linear SVM algorithm, we prepared a data set comprised of pairs of entangled and separable states.
The data set had 20000 samples of each family of states $\{\rm{G, E_{i}}, i=1,\dots 8\}$ and the same number of samples of separable states $S$.
Specifically, in the case of $\rm{G, E_{1} -E_{5}}$ we randomly sampled complex numbers $a,b,c,d$ describing parametrization of a particular family \eqref{G} - \eqref{E8}.
All states in the data set are normalized and converted to the density operator form as we mentioned above. 

It is important to emphasize that, in principle, an increase of the training data set size will improve the overall performance of any algorithm.
However, the computational resources required for such an improvement were not available for the full scope of this work.
The doubling of the training data set demonstrates at maximum $\approx 0.1 \%$ improvements of overall accuracy but roughly doubling computational time, plus one requires additional time to tune the algorithm's parameters.
Based on performance results we have chosen 20000 as a reasonable number of samples.
As a result, for each member of $\{\rm{E_{i}, i=1,\dots 8\}}$, we constructed a corresponding set of normalized EW operators such that $|{\rm{tr}}(\hat{\varrho}\hat{W})| \leq 1$.

\begin{figure*}
\centering
\includegraphics[width=0.95\textwidth]{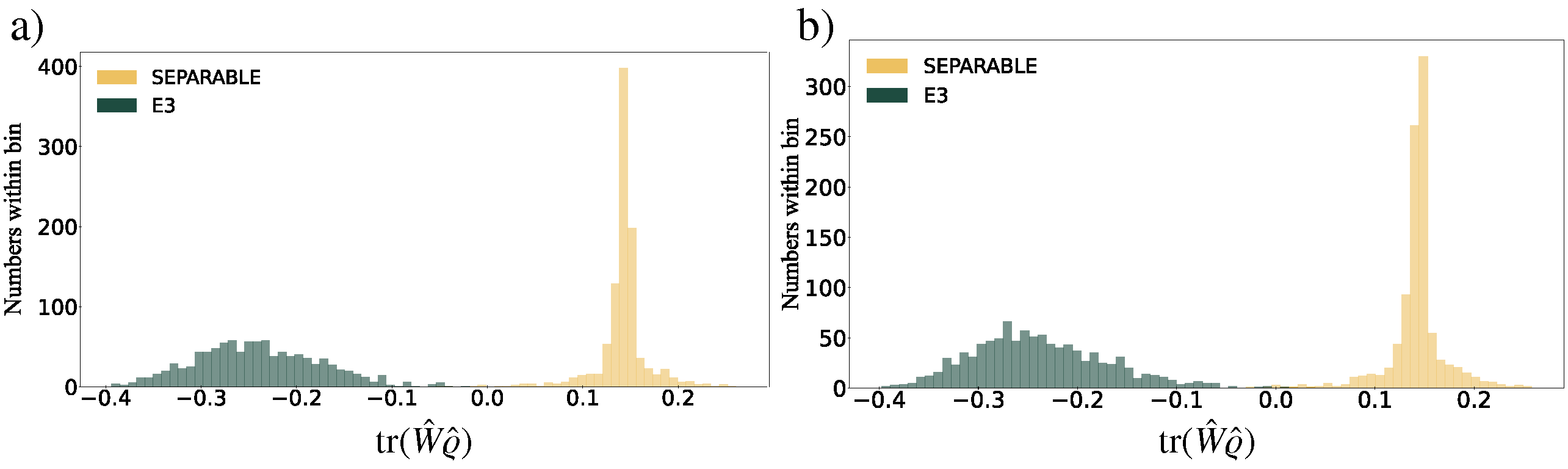}
\caption{Distribution of ${\rm{tr}}(\hat{\varrho}\hat{W})$, mean value of the entanglement witness for a) validation set and b) test set of the trained EW model via Linear SVM algorithm in the case of $\rm{{E}_{3}}$ family of states.
Both test set and validation set consist of 2000 samples: 1000 separable states and 1000 entangled states of $\rm{{E}_{3}}$ family.
For both validation and test sets there were only few ($\leq 5$) miss-classifications of entangled states and zero miss-classifications for separable states.
Note that the training data set included mixed Werner states to achieve better generalization.}
\label{RES_E3}
\end{figure*}

\begin{figure*}
\centering
\includegraphics[width=0.95\textwidth]{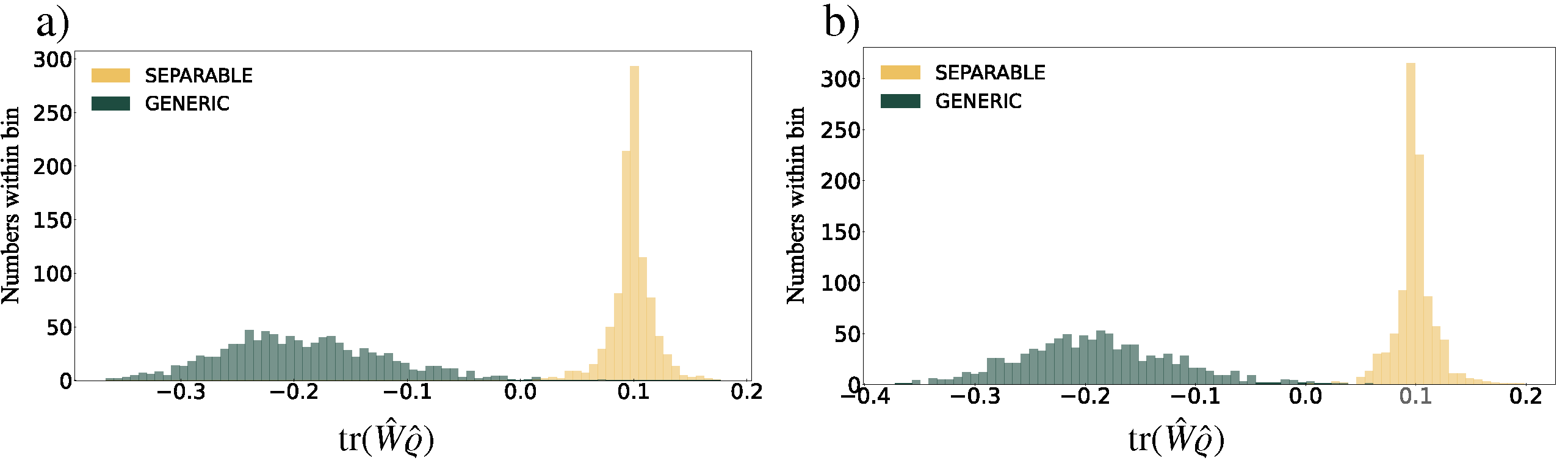}
\caption{Distribution of ${\rm{tr}}(\hat{\varrho}\hat{W})$, mean value of entanglement witness, for a) validation set and b) test set of trained EW model via Linear SVM algorithm in the case of generic $\rm{G}$ family of states. Both validation and test set consist of 2000 samples: 1000 separable states and 1000 entangled states of $\rm{{E}_{3}}$ family.
For both validation and test sets there were only few ($\leq 5$) miss-classifications of entangled states and zero miss-classifications for separable states.
Note that the training data set included mixed Werner states to achieve better generalization.}
\label{RES_GENERIC}
\end{figure*}

To train an SVM model for a given family, we split the corresponding data subset (for each pair of states) into three parts: 18000 samples were used for the training set, 1000 samples for {\it{ a validation set}} to tune hyperparameters and 1000 samples for {\it{a test set}} to check the generalization capabilities of an algorithm by evaluating the accuracy on another data set.
We presented the results of the trained model performance via SVM in Figures \ref{RES_E3} and \ref{RES_GENERIC} for the cases of families $\rm{E_{3}}$ and $\rm{G}$.
Histograms of  $\rm{E_{3}}$ and $\rm{G}$ consist of 75 bins versus ${\rm{tr}}(\hat{\varrho}\hat{W})$ covering the range of $|{\rm{tr}}(\hat{\varrho}\hat{W}_{G})|\leq 1$.
Note that for the both presented cases we have chosen the parameter $\varepsilon = 0.5$ in \eqref{U_sampling} to assemble data sets of states.
We set the regularization parameter in \eqref{hinge_loss},$ \lambda = 0.5 * 10^{-4}$ to yield an acceptable generalization, based on the performance on the validation data set.

We used the Adam optimizer implemented in {\textsc{Tensorflow}} library \cite{TF} with the following parameters: learning rate=0.005, epsilon=$10^{-6}$.
The total number of training steps, or epochs, was chosen to be equal to 20000 steps, but an acceptable convergence of SVM algorithm with the batch size 50 was achieved after approximately 5000 - 7000 training steps.
Based on the performance analysis on a validation set we utilized the following regularization strategy.
Each batch consisting of 50 samples was randomly sampled from the whole data set for each epoch.
This additional randomization strategy had demonstrated significant improvements on a test data set.
It is clear from the performance presented in Figure \ref{RES_E3} that the resulting witness can reliably distinguish separable and entangled states. 

Let us further analyze the performance of the linear SVM algorithm.
It is worth mentioning that, in accordance with the properties of multipartite EW \eqref{witness_general_def} $f_{W}(\varrho)\geq 0$ for all separable states but $f_{W}(\hat{\varrho}_{sep.})< 0$ for {\it{at least one}} entangled state $\hat{\varrho}_{ent.}$; it does not exist an EW operator that can detect all entangled states
We aim to find the best possible EW operator that can detect the presence of entanglement in a maximum number of states of a given family of states.

Usually, the optimal EW implies an operator $\hat{W}_{opt}$ that can detect the maximum number of entangled states \cite{Heinosaari2009}.
In geometrical terms it means that the decision hyperplane is tangent to the set of all separable states $S$.
Note that for a single entangled state it is guaranteed that this entangled state and all separable states can be linearly separated.
On the other hand, in terms of linear SVM we tend to find the maximal distance to the closest sample of a given set of states from the decision boundary.
It is known \cite{bishop2007}, that the linear SVM works perfectly for linearly separable data, but even for non linearly separable data it might perform quite well \cite{HandsOn}.
Below we will
demonstrate that analysis of the SVM performance can help to determine to what extent the entangled states from a given family can be linearly separated from a set of all separable states.

The analysis presented here directly tests the ability to distinguish entangled and separable states for each SLOCC family individually. The SVM generalization performance is considered for the newly resampled entangled and separable states.
In what follows, we analyze a variety of states correctly detected by EW without limiting them to only pure entangled states from families $\{\rm{G, E_{i}}, i=1,\dots 8\}$, but also considering a specific family of mixed entangled states - Werner states. Note that doing this we also included mixed Werner states and pure factorized states in the construction of a training data set. 

\begin{figure*}
\includegraphics[width=0.93\textwidth]{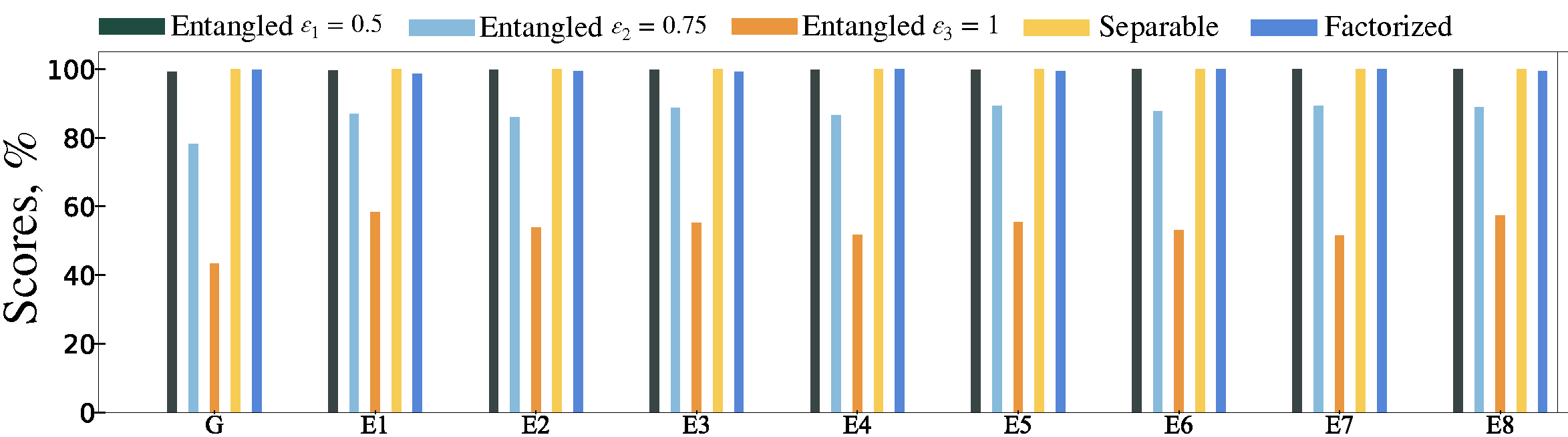}
\caption{General analysis of correct detection of entangled states based on trained set of EW operators.
For each family $\rm{G, E_{1}, \dots E_{8}}$ we used 20000 new samples of states.
Each state is modified according to Eq. \eqref{SLOCC_general}.
The random unitary operators are controlled by parameters $\varepsilon$ according to Eq. \eqref{U_sampling}.
The depicted bars represent the probability (or score) to correctly detect entangled states of a given family for various values of $\varepsilon$: 0.5, 0.75, 1, correspondingly for the bars 1, 2, 3 starting from the left.
For each family index ${\rm{G, E_{1} - E_{8}}}$ the fourth bar $4$ corresponds to  mixed separable states, the fifth bar $5$ corresponds to pure separable (factorized) states.
The set of EW is obtained employing SVM with parameters described in the caption of Figure (\ref{RES_E3}) and Fig.(\ref{RES_GENERIC}).}
\label{VER_RANDU}
\end{figure*}

Let us analyze the performance of a linear SVM classifier to train the entanglement witnesses corresponding to each family of quantum states.
We consider three different cases correspond to three parameters $\varepsilon_{1},\varepsilon_{2},\varepsilon_{3}$. In these cases we operate with different data sets of entangled states.
The diversity of each sample is varying by parameter $\varepsilon$  in \eqref{U_sampling}, in accordance with the SLOCC classification criteria \eqref{SLOCC_general}.
We tested the procedure for the following parameters: $\varepsilon_{1} = 0.5, \varepsilon_{2} = 0.75, \varepsilon_{3} = 1$.
The results are presented in Figure \ref{VER_RANDU}.
It is clear from the figure that the ability to detect entanglement with an already trained EW operator for the corresponding family members drops dramatically for $\varepsilon_{3} = 1$: the score is equal to about $50 \%$ on average for all classes compared to almost $100 \%$ for $\varepsilon_{1} = 0.5$.
In these cases, the value of $\varepsilon$ has a possible interpretation of ability to distinguish entangled states from separable ones employing {\it{linear}} SVM and, therefore, the EW theory  approach.

The approach works similarly for all families $\{E_{i}\}, i = 1\dots 8$.
In other words, one can roughly estimate the vicinity for any state if it is a member of a given family of entangled states, for which a linear model can detect the presence of entanglement. 
We also analyzed the SVM generalization performance in the worst case scenario. 
In this case, each state is modified via \eqref{SLOCC_general} but each unitary operator $\hat{U}_{f}^{(4)} \equiv \hat{U}_{1} \otimes \hat{U}_{2} \otimes \hat{U}_{3}\otimes \hat{U}_{4}$ is distributed according to the Haar measure.
As expected, in this case we have a much inferior  performance: scores $\leq 1 \% $ for all SLOCC families.
However, the correct detection scores for separable and factorized states have not changed and are still high, about $99.5 \% $.
Such behavior agrees with the EW properties. 

A pure factorized state always has an infinitely large number of entangled states within $\varepsilon$, the vicinity, calculated, for example, with respect to the trace norm or Hilbert-Schmidt norm.
This may lead to inferior performance of the SVM algorithm in its attempt to derive EWs capable of detecting the {\it{mixed entangled states}}.
Consequently, to improve the algorithm further we have included mixed entangled states in our training model.
For instance, let us consider another important family of multipartite quantum states $\hat{\varrho}_{W}$, so called Werner states \cite{Huber2021,Dariusz2006}, which involve entanglement of mixed states.

It is known \cite{Horodecki2009} that, in general, the detection of entanglement presence in Werner states utilizing entanglement witnesses is not an easy task.
Werner states are invariant under the diagonal action of the unitary group $U(2)\otimes U(2)$.
Thus, in the case of four-qubit quantum states the Werner states are defined as follows:
\begin{eqnarray}\label{WERNER_TRANSFORM}
\hat{\varrho}_{W} = [\hat{\mathbf{U}} \otimes\hat{\mathbf{U}}] \hat{\varrho}_{W}
[\hat{\mathbf{U}}^{\dagger}\otimes\hat{\mathbf{U}}^{\dagger}],
\end{eqnarray}
where $\hat{\mathbf{U}} = \hat{U}_{1}\otimes\hat{U}_{2}, \ U_{1},U_{2} \in U(2)$. 

To construct the sample data sets of both entangled and separable Werner states we followed derivations presented in  \cite{Dariusz2006}. 
An arbitrary Werner state can be obtained by action of a {\it{twril channel}}: $\hat{\varrho}_{W} = \tau\big[\hat{\varrho}_{W}\big] =  \int [\hat{\mathbf{U}} \otimes \hat{\mathbf{U}}] \hat{\varrho} [\hat{\mathbf{U}} \otimes \hat{\mathbf{U}}] d_{Haar} \mathbf{U} $ where $ d_{Haar} \mathbf{U} = dU_{1}dU_{2}$, invariant Haar measure on group $U(2)\otimes U(2)$.
This operation can be considered as a projection on $\hat{\mathbf{U}} \otimes \hat{\mathbf{U}}$, an invariant subspace. 

On the other hand, any Werner state can be represented as a decomposition $\hat{\varrho}_{W} = \sum_{\alpha}q_{\alpha}\hat{Q}_{\alpha}$, where $\{\hat{Q}_{\alpha}\}$ is a set of 4-partite orthogonal projectors that characterize $\hat{\mathbf{U}} \otimes \hat{\mathbf{U}}$ invariant subspace, see \cite{Dariusz2006} where an explicit form of the operators is presented).
It is evident that action of the twirling channel on a separable state results in a separable state.
Thus, as it was shown in \cite{Dariusz2006}, the  Werner state is separable if and only when the following conditions hold: $q_{1}\leq 1, q_{2},q_{3} \leq 1/2, q_{4}\leq 1/4; q_{4}\leq q_{2},q_{3}\leq 1$ for mentioned above decomposition of $\hat{\varrho}_{W}$.
In our analysis we applied this criteria to sample both separable and entangled states by considering the following procedures.
To sample the separable Werner states one can sample a simple four-qubit separable state and then apply a twirling channel.

Action of a twirling channel is equivalent to projection of a quantum state $\hat{\varrho}$ by using operators $\{\hat{Q}_{\alpha}\}$ with coefficients $q_{\alpha}(\hat{\varrho}) = {\rm{tr}}(\hat{Q}_{\alpha}\hat{\varrho}), \alpha = 1, \dots 4.$ to obtain corresponding $\hat{\varrho}_{W}(\hat{\varrho}) = \sum_{\alpha}q_{\alpha}(\hat{\varrho})\hat{Q}_{\alpha}$.
Such construction of Werner states works because the following properties of twirling channel hold. First, the dual channel to twirling channel is again a twirling channel. Second, any $\hat{Q}_{\alpha}$ is invariant under action of a twirling channel.
Entangled Werner states are obtained by sampling a random four - qubit state and collecting states for which the separability criteria had been violated.

Utilizing already trained witness operators for each SLOCC family we  found that an entanglement detection for the random entangled Werner states yielded poor accuracy. 
We obtained correct entanglement detection scores of just $\approx 1 \%$ based on a trained witness operator that corresponded to the family ${\rm{G}}$, for all other families of states we got scores ${\approx 0.1 \%}$.
On the other hand, the separable states are again classified with almost 100 \% accuracy.
Summarizing, almost all sampled entangled Werner states were classified as separable states.
Thus, it was essential to train the SVM linear modes for the Werner states and consider such states as a separate family.

\begin{figure*}
\centering
\includegraphics[width=0.92\textwidth]{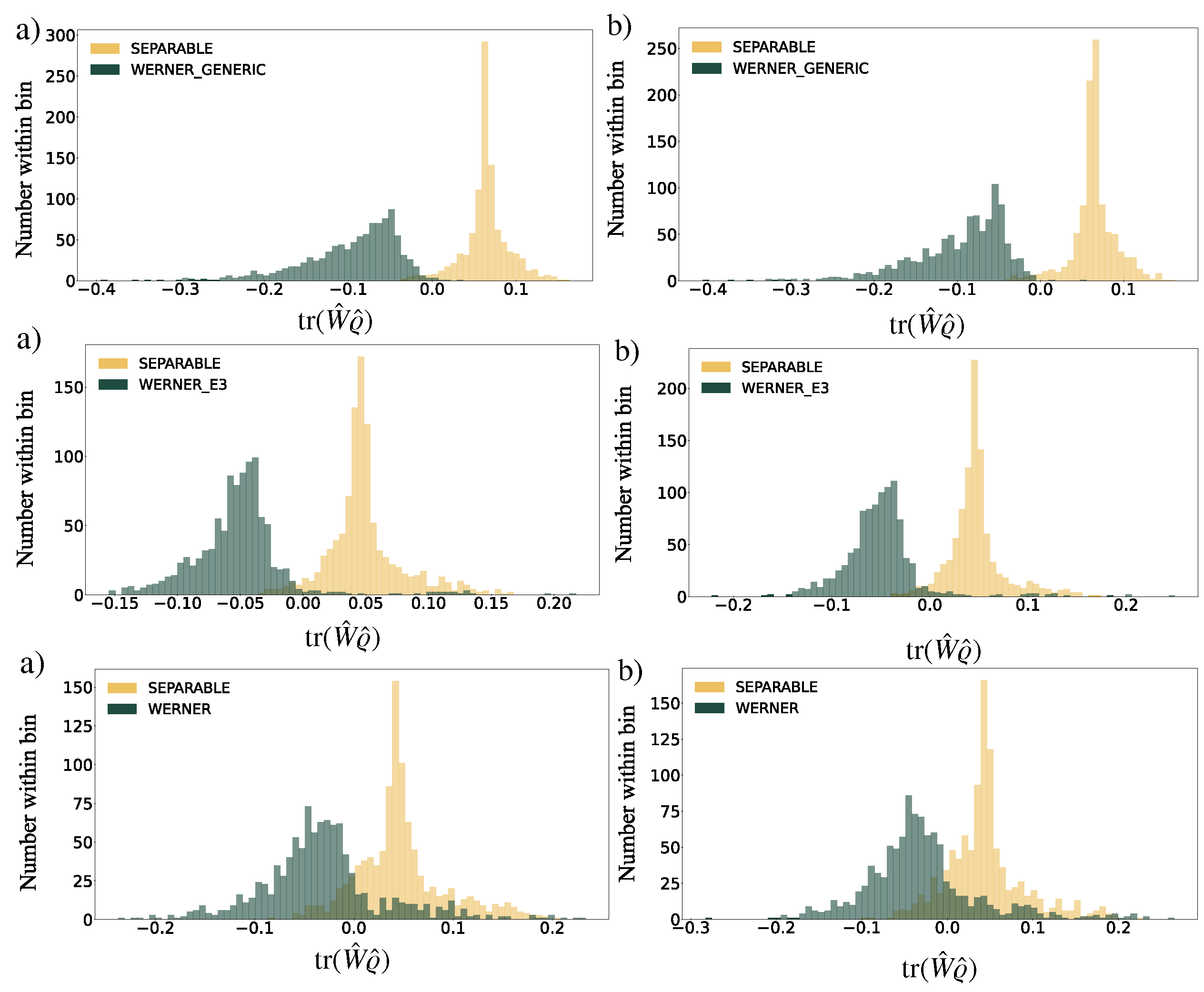}
\caption{Distribution of ${\rm{tr}}(\hat{\varrho}\hat{W})$, mean value of entanglement witness, for a) validation set and b) test set of trained EW model via Linear SVM algorithm in the case of Werner family of entangled states.
All parameters are the same as was described in Figure (\ref{RES_E3}) and Figure (\ref{RES_GENERIC}).
For the first and second rows in a plot the corresponding Werner states are generated from Generic-$\rm{G}$ and $\rm{E3}$ family of states by applying twirling channel and separability criteria, in a third row provides performance in the case of general sampling strategy described in a main text.}
\label{RES_WERNER}
\end{figure*}

The performance results of trained EW in the case of Werner states is presented in Figure \ref{RES_WERNER}.
It is not surprising that the  linear SVM performance was poor because the Werner states are not linearly separable in the space of trace class operators.
Nevertheless, it is clear from the separability criteria above (see also details in work \cite{Dariusz2006}) that the separable and entangled Werner states can be distinguished in invariant subspace $\hat{\mathbf{U}} \otimes \hat{\mathbf{U}}$.
Furthermore, one may project subsets of states parameterized with $\vec{\beta}$ from subsets ($S_{\vec{\beta}}$) of a given SLOCC family $\{\hat{\varrho}({\rm{E}}_{i},\vec{\beta}),\vec{\beta} \in S_{\vec{\beta}} \subseteq  \mathcal{R}^{m}\}$ onto invariant subspace $\hat{\mathbf{U}} \otimes \hat{\mathbf{U}}$. In a such case one may assume that even a linear SVM model can yield good performance.
Indeed, one may expect that knowing a  trained entanglement witness $\hat{W}(\vec{\beta}):{\rm{tr}}\left(\hat{W}(\vec{\beta})\hat{\varrho}(\vec{\beta})\right)<0,\forall \vec{\beta} \in S_{\vec{\beta}}$ one may construct an EW operator for the corresponding subset of Werner states obtained via a twirling channel $\hat{\varrho}_{W}(\vec{\beta}) =\tau \big[\hat{\varrho}(\vec{\beta})\big]$.
Mathematically, it is evident from the following example. By choosing a family $\hat{\varrho}(\vec{\beta})\in\rm{G, E_{i}, i = 1, \dots 8}, \forall  \vec{\beta} \in  S_{\vec{\beta}} $, and assuming that  there  exist an EW operator $\hat{W}'(\vec{\beta})$ such that ${\rm{sign}}\big[{\rm{tr}}(\tau\big[\hat{\varrho}(\vec{\beta})\big]\hat{W}'(\vec{\beta}))\big] = {\rm{sign}}\big[{\rm{tr}}(\hat{\varrho}(\vec{\beta})\tau\big[\hat{W'}(\vec{\beta})\big])\big] < 0 \ \forall \vec{\beta} \in S_{\vec{\beta}}$. In the  above equality we utilized the self-duality property of a twirling channel by swapping action of $\tau\big[\hat{\varrho}(\vec{\beta})\big] \longrightarrow \tau\big[\hat{W'}(\vec{\beta})\big])\big] $ under the trace operation. Thus, one may imply that
$\tau\big[\hat{W'}(\vec{\beta})\big] \subset {\rm{span}}(\{\hat{P}_{\lambda_{j}}(\hat{W}(\vec{\beta}))\})$.
 We denoted $"\rm{span}(\cdot)"$ as linear span for a set of operators.
 
In our case these sets of operators $\{\hat{P}_{\lambda_{j}}(\hat{W}(\vec{\beta}))\}$ are orthogonal projectors on an eigenspace that corresponds to an eigenvalue $\lambda_{j} \in {\rm{spec}}(\hat{W}(\vec{\beta}))$.
It is evident that if the set $S_{\vec{\beta}}$ is simple enough the correspondence can be specified straightforwardly.
As we can see from Figure \ref{RES_WERNER} the direct numerical evaluation of linear SVM algorithm support the discussed assumption.
Note, an approach discussed above might be generalized further to construct more sophisticated algorithm for entanglement detection of multipartite quantum states.

In this subsection, we considered several families of entangled states, including entangled Werner states.
Our analysis clearly indicates that a linear SVM algorithm can be efficiently employed to construct a set of EW operators.
However, it was also demonstrated that the approach has fundamental limitations. 
It is evident because each family of states is quite complex and cannot be separated by applying just a linear SVM algorithm.
In the following, we discuss possible nonlinear extensions of the SVM algorithms applicable to the problem of entanglement detection.

\subsection{Constructing  Entanglement Witness operators via Support Vector Machines with nonlinear kernels}\label{SVM_kernel}

Let us consider possible improvements of the discussed above approach.
The SVM method is a quadratic constrained programming problem, which can be seen from Equation \eqref{hinge_loss}.
Importantly, the maximum margin problem has a dual representation operating with kernel functions \cite{bishop2007,HandsOn}.
We can upgrade our simple linear model \eqref{svm_simpliest_decision_function} with the following expression:
\begin{eqnarray}\label{kermap}
f_{\vec{\phi}}(\vec{\varrho}) = \left(\vec{w}(\vec{\varrho}), \  \vec{\phi}(\vec{\varrho})\right),
\end{eqnarray}
where a map $\vec{\phi}(\cdot)$ transforms the initial feature space, but the overall model is still linear.
One can write the error function (loss) in the following form:

\begin{equation}\label{dualerror}
\Tilde{L}(\vec{a}) = \sum_{i}^{N}a_{n} - \frac{1}{2}\sum_{n,m}^{N}a_{n}a_{m}t_{n}t_{m}k(\vec{\varrho}_{n}, \vec{\varrho}_{m}),
\end{equation}
where we introduced a new vector $\vec{a} = (a_{1}, \dots a_{N})$ related to  $\vec{w}(\vec{\varrho})$ as follows: $\vec{w}(\vec{\varrho}) = \sum_{i}^{N}a_{n}t_{n}\vec{\phi}(\vec{\varrho})$.
The symbol $k(\vec{\varrho}_{n}, \vec{\varrho}_{m}$) represents a {\it{kernel function}}, which can be efficiently used to evaluate nonlinear transformations.
The optimization (training by finding $\vec{a}$) of quadratic function \eqref{dualerror} yields the following predictive model in terms of kernel function:
\begin{equation}\label{nonlinearmodel}
f_{\vec{\phi}}(\vec{\varrho}) \longrightarrow f_{k(\vec{\varrho}_{n},\cdot)}(\vec{\varrho}) = \sum_{i}^{N}a_{n}t_{n}k(\vec{\varrho}_{n},\vec{\varrho})
\end{equation}

\begin{figure*}
\centering
\includegraphics[width=0.95\textwidth]{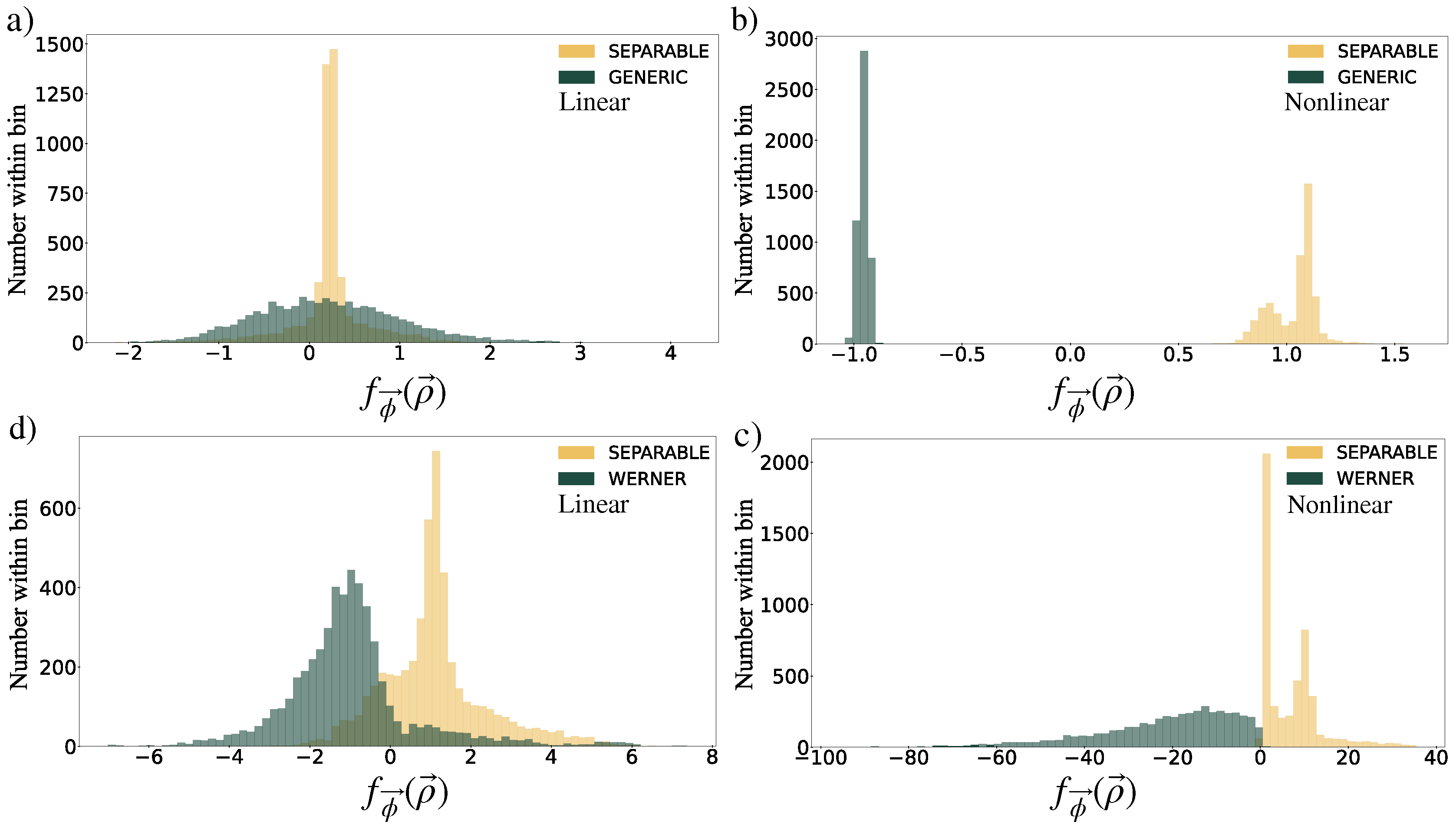}
\caption{Distribution of $f_{\vec{\phi}}(\vec{\varrho})$ described in Eqs. \eqref{kermap} and \eqref{nonlinearmodel}, and comparison of trained linear and nonlinear SVM algorithm with RBF kernel using SciKit Learn Python package. 
Cases a) and d) correspond to the standard SVM classification with linear kernel on a newly generated test set, whereas b) and c) correspond to the Generic and entangled Werner states classified with a nonlinear SVM with RBF kernel.
We verify results by sampling 5000 states of each family and for each case. 
Note that the assembled data set of Generic states is obtained by modifying each sample by unitary operators distributed according to the Haar measure and Eq.\ref{SLOCC_general}. 
It is clear that the performance of nonlinear SVM is very good, 100\% and 99,8\% in the cases of Generic and Werner states, respectively.}
\label{NONLINEAR}
\end{figure*}

Note that working with kernel functions allows one to avoid explicit operations with the featured space, though it could produce, in principle, more successful and complex models.
For instance, it could happen that one finds a specific kernel that can perfectly distinguish separable states and the {\it{whole}} family of entangled states based on the SLOCC classification.
Indeed, we verified this assumption directly applying a nonlinear SVM using {\textsc{SciKit Learn}} package.
Figure (\ref{NONLINEAR}) provides a comparison of models with linear and nonlinear kernels to classify the entangled Werner, generic and separable states.
Furthermore, we modify the generic states according to the Haar measure to assemble a new data set of 1000 samples. 
We observe 100 \% accuracy on a test set in the case of nonlinear SVM with Radial Basis Function (RBF) kernel: $k(\vec{\varrho},\vec{\sigma}) = {\rm{exp}}\left(-\gamma||\vec{\varrho} - \vec{\sigma}||_{2}^{2}\right)$, compared to the linear model.

Unfortunately, there are constraints among the feature values that restrict the dimension of feature space so evaluating kernel functions for all samples might be computationally demanding. 
To reduce the problem complexity, one needs to find certain heuristics or another advanced coarse-grained classification.
In the next section we summarize the obtained results and discuss possible applications and further steps.

\section{Discussion}

Let us summarize the salient points regarding the problem of entanglement detection in multipartite states using the support vector machine algorithms as we considered in our work. 

The entanglement detection problem is rather complex.
Ideally, one would aim to find tangent hyperplanes at each point of a set of separable states. 
Several methods were proposed to find an approximate distance between a given witness and a set of separable states, which help to detect the entanglement of a particular state \cite{Ioannou2007}.
In this context the considered SVM approach has a similar objective to the one mentioned above. 
The SVM aims to maximize the shortest distance between the decision boundary and "support" state (vector). However, as it was shown in \cite{Ioannou2004,Ioannou2006,Ioannou2007} the characterization of all EWs for a given multipartite system is a NP-hard problem as the dimensionality increases. 
The geometrical interpretation of this problem is also a highly nontrivial task \cite{bengtsson2017geometry}.  

Nevertheless, the EW problem complexity can be reduced by employing a coarse-grained classification such as the considered four-qubit SLOCC classification or description in terms of Werner states.
We demonstrated successful applications of the linear SVM algorithm for both of these cases.
It is important to mention that our analysis is akin to that proposed in \cite{zhu2021}, where the authors also emphasized the direct relationship of the entanglement witness functional and linear SVM. 
In our work here, we focus on the successful applicability of the SVM technique specifically for the coarse-grained classification of entangled states joined in to certain families.
However, the success of linear models comes with a price owing to the mentioned above NP-hardness of the EW problem.
In general, the linear SVM approach cannot provide high accuracy for all states in a particular family.
As we showed, one can detect the entanglement with high accuracy for all family members only within a vicinity of a fiducial state of the family. 
Thus, one can interpret the accuracy of a trained EW model based on linear SVM as a measure of complexity for a given family or parametrization of entangled states. 

One of the main advantages of the discussed linear SVM approach is its universality in terms of implementation.
Note that the modern Python packages such as {\textsc{QuTip,Tensorflow, Scikit learn}}\cite{QUTIP2012,QUTIP2013,tensorflow2015-whitepaper,pedregosa2011scikit} provide all necessary tools for implementation of both linear and nonlinear SVM to find EW. The source code used for this work is also provided in a GitHub repository \cite{github}.
In addition, a nonlinear version of SVM can be a valuable tool for theoretical analysis, especially in the case of coarse-grained classifications.
We demonstrated that nonlinear SVM allows to detect the entanglement with almost perfect accuracy for any arbitrary state from a given family, while the linear SVM has almost zero accuracy.
It is not surprising because an arbitrary family of states based on a given classification or parametrization is highly likely to be not linearly separable from a convex set of separable states.
We also showed that introduction of the kernel function could be useful for specific cases.
At the same time it may require considerable computing resources making the nonlinear kernels less attractive in general. 

It is important to mention that local unitary transformations limit our analysis.
These limitations are dictated by the simplicity of analysis and implementation of the SVM algorithm. Nevertheless, we believe that our results are sufficient to demonstrate the universality of the proposed approach.
It is also worth noting that the previously described problematic Werner states are not members of the SLOCC family, as their definition is based on the unitary group structure described above.
In addition, we tailored our work to specific practical problems, such as quantum astrometry \cite{Stankus2022, LUKIN1,LUKIN2}, where the local unitary operations are reasonable models for the information encoding present in those environments. 
On the other hand, it is also crucial to emphasize that the general analysis of the symmetry of quantum states and operations, for example, by applying group theory, plays a prominent role in the construction of the coarse-grained description.

Application of the group theory and symmetry analysis are some of the most powerful tools in theoretical physics. 
In particular, these approaches led to a significant progress in developing novel theoretical and practical tools for quantum technologies and their applications, such as entanglement detection, foundations of quantum theory \cite{Graydon2016,Graydon2017conical,Graydon2021,Graydon2022}, randomized benchmarking \cite{Erhard2019,Harper2020,Joel2021} etc.
The aim for the next stage of research is to elaborate on the developments of SVM - like algorithms to train more sophisticated and, if possible, universal frameworks of classifiers to combine specific sets of multipartite quantum states.
It is our hope that the synthesis of machine learning-based group theory methods and symmetry analysis of states and quantum channels can be valuable tools to accelerate further developments of quantum technology.

Another advantage of linear models is that they allow for a physical interpretation of EW operators, as observables that can be measured directly in an experiment.
An interesting goal would be to analyze further the connection between kernel methods and optimal collective measurements within the concept of collective entanglement witnesses \cite{Horodecki2003b}.

The presented SVM-based analysis of EW can be useful for practical applications.
For instance, the SLOCC operations described above directly relate to the field of quantum metrology.
In particular, optical interferometers were proposed where an entangled ancilla was shared between two or more stations to improve the accuracy of astrometrical measurements \cite{Gottesman,LUKIN1,LUKIN2,Stankus2022,INSTR}.
Each ancilla's subsystem interacts locally with a fiducial state in the station that carries valuable information to be extracted. 
For practical implementation of such schemes, one needs to understand how the noise affects the ancilla's state and also specifics of local interactions to optimize the measurement protocol and to quantify the structure of the entangled states. 
Simple and effective tools such as the proposed SVM approach can help to solve the problem of detecting and classifying the structure of entangled states used in those quantum astrometry schemes.

Finally, another promising application of this approach is its employment in quantum-enhanced sensor networks\cite{Zhuang2019} and quantum reservoir computing \cite{Ghosh2019,vintskevich2022computing}. 
It was already demonstrated that both SVM and EW problems could be mapped and processed by such multipartite quantum systems working as quantum-enhanced processors \cite{Zhuang2019}. 
These promising developments open new directions for applications of multipartite quantum systems to address the cross-cutting interdisciplinary problems of machine learning and quantum metrology.

\section{Acknowledgments}
We are grateful to Matthew Graydon, Zhi Chen, An\v{z}e Slosar, Konstantin Katamadze and Rene Reimann for useful discussions and support. This work was supported in part by the U.S. Department of Energy QuantISED award and BNL LDRD grant 19-30.

\appendix
\section{SLOCC classification of four-qubit pure states}\label{appendix-a}
All trained SVM models discussed above were derived by combining all states of one family based on a given property such as the SLOCC equivalence or another particular state property (parametrization) intrinsic to the Werner states. 
This appendix provides an explicit description of the SLOCC classification used in this manuscript. 
This classification was investigated in detail in  \cite{Verstraete2002}, where the authors derived nine families of states corresponding to nine different ways of entangling four qubits.

The main idea of derivation follows from the equivalence of groups ${\rm{SU}}(2)\otimes{\rm{SU}}(2)$ and ${\rm{SO}}(4)$ in the Lie-group theory, where ${\rm{SU}}(2)$ is a special group of $2{\rm{x}}2$ unitary matrices and ${\rm{SO}}(4)$ is a group of orthogonal matrices with unit determinant. 
Mathematically, this equivalence is represented by $\forall U_{1},U_{2} \in {\rm{SU}}(2) \ \exists T: O = T(U_{1}\otimes U_{2})T^{\dagger} \in  {\rm{SO}}(4)$.
Additionally, each four-qubit pure state can be represented as a $4{\rm{x}}4$ complex matrix $R =T\psi_{(i_{1}i_{2})(i_{3}i_{4})}T^{\dagger}$, where $\psi_{(i_{1}i_{2})(i_{3}i_{4})}$ is a reshaped matrix element of four-qubit state representation in the computational basis. 
Thus, the aforementioned equivalence and state representation allows one to represent the SLOCC transformation of four-qubit state in \eqref{SLOCC_general} as an action of orthogonal matrices: $R' = O_{1}RO_{2}$. 
On the other hand it was proven that a given $4{\rm{x}}4$ complex matrix $R$ can always be transformed via $O_{1}$ and $O_{2}$ to the Jordan block normal form. 
This means that the normal form encodes the genuine non-local properties of the four-qubit state and allows classification of the states. 
Entangled states from the same family can perform the same quantum protocol but with a different probability. 
The following distinct classes of (unnormalized) pure states were specified in the computational basis:
\begin{eqnarray}\label{G}
&&\ket{G_{abcd}} =\nonumber\\
&&\frac{a+d}{2}(\ket{0000} + (\ket{1111}) + \frac{a-d}{2}(\ket{0011} + \ket{1100}) + \nonumber \\
&&\frac{b+c}{2}(\ket{0101} +\ket{1010}) + \nonumber \\
&&+\frac{b-c}{2}(\ket{0110} + \ket{1001})
\end{eqnarray}

\begin{eqnarray}\label{E1}
&&\ket{E_{1}} = \frac{a+b}{2}(\ket{0000} + \ket{1111})+\frac{a-b}{2}(\ket{0011} \nonumber \\
&& + \ket{1100}) +c(\ket{0101} + \ket{1010}) + \ket{0110}
\end{eqnarray}

\begin{eqnarray}\label{E2}
&&\ket{E_{2}}=a(\ket{0000}+\ket{1111}) + b(\ket{0101}+\ket{1010}) \nonumber \\
&& +\ket{0110} + \ket{0011}
\end{eqnarray}

\begin{eqnarray}\label{E3}
&&\ket{E_{3}} = a(\ket{0000} + \ket{1111}) + \frac{a + b}{2}(\ket{0101} + \ket{1010})  \nonumber\\ 
&&+\frac{a-b}{2}(\ket{0110}+\ket{1001}) \nonumber \\
&&+\frac{i}{\sqrt{2}}(\ket{0001} + \ket{0010}+\ket{0111} + \ket{1011})
\end{eqnarray}

\begin{eqnarray}\label{E4}
&&\ket{E_{4}} = a(\ket{0000} + \ket{0101}) + \ket{1010} + \ket{1111}) \nonumber \\
&& + (i\ket{0001} + \ket{0110} - i\ket{1011})
\end{eqnarray}

\begin{eqnarray}\label{E5}
&&\ket{E_{5}} = a(\ket{0000} + \ket{1111}) \nonumber\\
&&+ (\ket{0011} + \ket{0101} + \ket{0110})
\end{eqnarray}

\begin{eqnarray}\label{E6}
\ket{E_{6}} = \ket{0000} + \ket{0101} + \ket{1000} + \ket{1110}
\end{eqnarray}
\begin{eqnarray}\label{E7}
\ket{E_{7}} = \ket{0000} + \ket{1011} + \ket{1101} + \ket{1110}
\end{eqnarray}
\begin{eqnarray}\label{E8}
\ket{E_{8}} = \ket{0000} + \ket{0111}
\end{eqnarray}
\begin{eqnarray}
\ket{S_{0}} = \ket{0110} \equiv \ket{E}_{1}:a=b=c=0
\end{eqnarray}

The state $\ket{S_{0}}$ represents a subclass of the pure factorized state.
All other states (classes) possess unique properties and entanglement structure. 
For instance, the state $\ket{G}_{abcd}$ represents a class of generic pure states.
It is claimed that this is a class of states with maximal 4-partite entanglement on the orbit generated by \rm{SLOCC} measured in accordance with majorization criteria \cite{Verstraete2002}.
\newpage

\begin{thebibliography}{66}
\expandafter\ifx\csname natexlab\endcsname\relax\def\natexlab#1{#1}\fi
\expandafter\ifx\csname bibnamefont\endcsname\relax
  \def\bibnamefont#1{#1}\fi
\expandafter\ifx\csname bibfnamefont\endcsname\relax
  \def\bibfnamefont#1{#1}\fi
\expandafter\ifx\csname citenamefont\endcsname\relax
  \def\citenamefont#1{#1}\fi
\expandafter\ifx\csname url\endcsname\relax
  \def\url#1{\texttt{#1}}\fi
\expandafter\ifx\csname urlprefix\endcsname\relax\def\urlprefix{URL }\fi
\providecommand{\bibinfo}[2]{#2}
\providecommand{\eprint}[2][]{\url{#2}}

\bibitem[{\citenamefont{Horodecki et~al.}(2009)\citenamefont{Horodecki,
  Horodecki, Horodecki, and Horodecki}}]{Horodecki2009}
\bibinfo{author}{\bibfnamefont{R.}~\bibnamefont{Horodecki}},
  \bibinfo{author}{\bibfnamefont{P.}~\bibnamefont{Horodecki}},
  \bibinfo{author}{\bibfnamefont{M.}~\bibnamefont{Horodecki}},
  \bibnamefont{and}
  \bibinfo{author}{\bibfnamefont{K.}~\bibnamefont{Horodecki}},
  \bibinfo{journal}{Rev. Mod. Phys.} \textbf{\bibinfo{volume}{81}},
  \bibinfo{pages}{865} (\bibinfo{year}{2009}),
  \urlprefix\url{https://link.aps.org/doi/10.1103/RevModPhys.81.865}.

\bibitem[{\citenamefont{Jozsa and Linden}(2003)}]{Jozsa2003}
\bibinfo{author}{\bibfnamefont{R.}~\bibnamefont{Jozsa}} \bibnamefont{and}
  \bibinfo{author}{\bibfnamefont{N.}~\bibnamefont{Linden}},
  \bibinfo{journal}{Proceedings of the Royal Society of London. Series A:
  Mathematical, Physical and Engineering Sciences}
  \textbf{\bibinfo{volume}{459}}, \bibinfo{pages}{2011} (\bibinfo{year}{2003}),
  \eprint{https://royalsocietypublishing.org/doi/pdf/10.1098/rspa.2002.1097},
  \urlprefix\url{https://royalsocietypublishing.org/doi/abs/10.1098/rspa.2002.1097}.

\bibitem[{\citenamefont{Briegel and Raussendorf}(2001)}]{Cluster2001}
\bibinfo{author}{\bibfnamefont{H.~J.} \bibnamefont{Briegel}} \bibnamefont{and}
  \bibinfo{author}{\bibfnamefont{R.}~\bibnamefont{Raussendorf}},
  \bibinfo{journal}{Phys. Rev. Lett.} \textbf{\bibinfo{volume}{86}},
  \bibinfo{pages}{910} (\bibinfo{year}{2001}),
  \urlprefix\url{https://link.aps.org/doi/10.1103/PhysRevLett.86.910}.

\bibitem[{\citenamefont{Zhang et~al.}(2006)\citenamefont{Zhang, Lu, Zhou, Chen,
  Zhao, Yang, and Pan}}]{Zhang2006}
\bibinfo{author}{\bibfnamefont{A.-N.} \bibnamefont{Zhang}},
  \bibinfo{author}{\bibfnamefont{C.-Y.} \bibnamefont{Lu}},
  \bibinfo{author}{\bibfnamefont{X.-Q.} \bibnamefont{Zhou}},
  \bibinfo{author}{\bibfnamefont{Y.-A.} \bibnamefont{Chen}},
  \bibinfo{author}{\bibfnamefont{Z.}~\bibnamefont{Zhao}},
  \bibinfo{author}{\bibfnamefont{T.}~\bibnamefont{Yang}}, \bibnamefont{and}
  \bibinfo{author}{\bibfnamefont{J.-W.} \bibnamefont{Pan}},
  \bibinfo{journal}{Phys. Rev. A} \textbf{\bibinfo{volume}{73}},
  \bibinfo{pages}{022330} (\bibinfo{year}{2006}),
  \urlprefix\url{https://link.aps.org/doi/10.1103/PhysRevA.73.022330}.

\bibitem[{\citenamefont{Wei et~al.}(2022)\citenamefont{Wei, Jing, Zhang, Liao,
  Yuan, Fan, Lyu, Zhou, Wang, Deng et~al.}}]{Wei2022}
\bibinfo{author}{\bibfnamefont{S.-H.} \bibnamefont{Wei}},
  \bibinfo{author}{\bibfnamefont{B.}~\bibnamefont{Jing}},
  \bibinfo{author}{\bibfnamefont{X.-Y.} \bibnamefont{Zhang}},
  \bibinfo{author}{\bibfnamefont{J.-Y.} \bibnamefont{Liao}},
  \bibinfo{author}{\bibfnamefont{C.-Z.} \bibnamefont{Yuan}},
  \bibinfo{author}{\bibfnamefont{B.-Y.} \bibnamefont{Fan}},
  \bibinfo{author}{\bibfnamefont{C.}~\bibnamefont{Lyu}},
  \bibinfo{author}{\bibfnamefont{D.-L.} \bibnamefont{Zhou}},
  \bibinfo{author}{\bibfnamefont{Y.}~\bibnamefont{Wang}},
  \bibinfo{author}{\bibfnamefont{G.-W.} \bibnamefont{Deng}},
  \bibnamefont{et~al.}, \bibinfo{journal}{Laser {\&} Photonics Reviews}
  \textbf{\bibinfo{volume}{16}}, \bibinfo{pages}{2100219}
  (\bibinfo{year}{2022}),
  \urlprefix\url{https://doi.org/10.1002/lpor.202100219}.

\bibitem[{\citenamefont{Hu et~al.}(2021)\citenamefont{Hu, Huang, Sheng, Zhou,
  Liu, Guo, Zhang, Xing, Huang, Li et~al.}}]{Hu2021}
\bibinfo{author}{\bibfnamefont{X.-M.} \bibnamefont{Hu}},
  \bibinfo{author}{\bibfnamefont{C.-X.} \bibnamefont{Huang}},
  \bibinfo{author}{\bibfnamefont{Y.-B.} \bibnamefont{Sheng}},
  \bibinfo{author}{\bibfnamefont{L.}~\bibnamefont{Zhou}},
  \bibinfo{author}{\bibfnamefont{B.-H.} \bibnamefont{Liu}},
  \bibinfo{author}{\bibfnamefont{Y.}~\bibnamefont{Guo}},
  \bibinfo{author}{\bibfnamefont{C.}~\bibnamefont{Zhang}},
  \bibinfo{author}{\bibfnamefont{W.-B.} \bibnamefont{Xing}},
  \bibinfo{author}{\bibfnamefont{Y.-F.} \bibnamefont{Huang}},
  \bibinfo{author}{\bibfnamefont{C.-F.} \bibnamefont{Li}},
  \bibnamefont{et~al.}, \bibinfo{journal}{Phys. Rev. Lett.}
  \textbf{\bibinfo{volume}{126}}, \bibinfo{pages}{010503}
  (\bibinfo{year}{2021}),
  \urlprefix\url{https://link.aps.org/doi/10.1103/PhysRevLett.126.010503}.

\bibitem[{\citenamefont{Lloyd}(2003)}]{Lloyd2003}
\bibinfo{author}{\bibfnamefont{S.}~\bibnamefont{Lloyd}},
  \bibinfo{journal}{Phys. Rev. Lett.} \textbf{\bibinfo{volume}{90}},
  \bibinfo{pages}{167902} (\bibinfo{year}{2003}),
  \urlprefix\url{https://link.aps.org/doi/10.1103/PhysRevLett.90.167902}.

\bibitem[{\citenamefont{McCutcheon et~al.}(2016)\citenamefont{McCutcheon,
  Pappa, Bell, McMillan, Chailloux, Lawson, Mafu, Markham, Diamanti, Kerenidis
  et~al.}}]{McCutcheon2016}
\bibinfo{author}{\bibfnamefont{W.}~\bibnamefont{McCutcheon}},
  \bibinfo{author}{\bibfnamefont{A.}~\bibnamefont{Pappa}},
  \bibinfo{author}{\bibfnamefont{B.~A.} \bibnamefont{Bell}},
  \bibinfo{author}{\bibfnamefont{A.}~\bibnamefont{McMillan}},
  \bibinfo{author}{\bibfnamefont{A.}~\bibnamefont{Chailloux}},
  \bibinfo{author}{\bibfnamefont{T.}~\bibnamefont{Lawson}},
  \bibinfo{author}{\bibfnamefont{M.}~\bibnamefont{Mafu}},
  \bibinfo{author}{\bibfnamefont{D.}~\bibnamefont{Markham}},
  \bibinfo{author}{\bibfnamefont{E.}~\bibnamefont{Diamanti}},
  \bibinfo{author}{\bibfnamefont{I.}~\bibnamefont{Kerenidis}},
  \bibnamefont{et~al.}, \bibinfo{journal}{Nature Communications}
  \textbf{\bibinfo{volume}{7}} (\bibinfo{year}{2016}),
  \urlprefix\url{https://doi.org/10.1038/ncomms13251}.

\bibitem[{\citenamefont{Illiano
  et~al.}(2022{\natexlab{a}})\citenamefont{Illiano, Caleffi, Manzalini, and
  Cacciapuoti}}]{Illiano2022}
\bibinfo{author}{\bibfnamefont{J.}~\bibnamefont{Illiano}},
  \bibinfo{author}{\bibfnamefont{M.}~\bibnamefont{Caleffi}},
  \bibinfo{author}{\bibfnamefont{A.}~\bibnamefont{Manzalini}},
  \bibnamefont{and} \bibinfo{author}{\bibfnamefont{A.~S.}
  \bibnamefont{Cacciapuoti}}, \bibinfo{journal}{Computer Networks}
  \textbf{\bibinfo{volume}{213}}, \bibinfo{pages}{109092}
  (\bibinfo{year}{2022}{\natexlab{a}}),
  \urlprefix\url{https://doi.org/10.1016/j.comnet.2022.109092}.

\bibitem[{\citenamefont{Illiano
  et~al.}(2022{\natexlab{b}})\citenamefont{Illiano, Viscardi, Koudia, Caleffi,
  and Cacciapuoti}}]{Jessica2022}
\bibinfo{author}{\bibfnamefont{J.}~\bibnamefont{Illiano}},
  \bibinfo{author}{\bibfnamefont{M.}~\bibnamefont{Viscardi}},
  \bibinfo{author}{\bibfnamefont{S.}~\bibnamefont{Koudia}},
  \bibinfo{author}{\bibfnamefont{M.}~\bibnamefont{Caleffi}}, \bibnamefont{and}
  \bibinfo{author}{\bibfnamefont{A.~S.} \bibnamefont{Cacciapuoti}},
  \emph{\bibinfo{title}{Quantum internet: from medium access control to
  entanglement access control}} (\bibinfo{year}{2022}{\natexlab{b}}),
  \urlprefix\url{https://arxiv.org/abs/2205.11923}.

\bibitem[{\citenamefont{T\'oth}(2012)}]{Toth2012}
\bibinfo{author}{\bibfnamefont{G.}~\bibnamefont{T\'oth}},
  \bibinfo{journal}{Phys. Rev. A} \textbf{\bibinfo{volume}{85}},
  \bibinfo{pages}{022322} (\bibinfo{year}{2012}),
  \urlprefix\url{https://link.aps.org/doi/10.1103/PhysRevA.85.022322}.

\bibitem[{\citenamefont{Hyllus et~al.}(2012)\citenamefont{Hyllus, Laskowski,
  Krischek, Schwemmer, Wieczorek, Weinfurter, Pezz\'e, and
  Smerzi}}]{Hyllus2012}
\bibinfo{author}{\bibfnamefont{P.}~\bibnamefont{Hyllus}},
  \bibinfo{author}{\bibfnamefont{W.}~\bibnamefont{Laskowski}},
  \bibinfo{author}{\bibfnamefont{R.}~\bibnamefont{Krischek}},
  \bibinfo{author}{\bibfnamefont{C.}~\bibnamefont{Schwemmer}},
  \bibinfo{author}{\bibfnamefont{W.}~\bibnamefont{Wieczorek}},
  \bibinfo{author}{\bibfnamefont{H.}~\bibnamefont{Weinfurter}},
  \bibinfo{author}{\bibfnamefont{L.}~\bibnamefont{Pezz\'e}}, \bibnamefont{and}
  \bibinfo{author}{\bibfnamefont{A.}~\bibnamefont{Smerzi}},
  \bibinfo{journal}{Phys. Rev. A} \textbf{\bibinfo{volume}{85}},
  \bibinfo{pages}{022321} (\bibinfo{year}{2012}),
  \urlprefix\url{https://link.aps.org/doi/10.1103/PhysRevA.85.022321}.

\bibitem[{\citenamefont{Gross et~al.}(2009)\citenamefont{Gross, Flammia, and
  Eisert}}]{Gross2009}
\bibinfo{author}{\bibfnamefont{D.}~\bibnamefont{Gross}},
  \bibinfo{author}{\bibfnamefont{S.~T.} \bibnamefont{Flammia}},
  \bibnamefont{and} \bibinfo{author}{\bibfnamefont{J.}~\bibnamefont{Eisert}},
  \bibinfo{journal}{Phys. Rev. Lett.} \textbf{\bibinfo{volume}{102}},
  \bibinfo{pages}{190501} (\bibinfo{year}{2009}),
  \urlprefix\url{https://link.aps.org/doi/10.1103/PhysRevLett.102.190501}.

\bibitem[{\citenamefont{D\"{u}r and Briegel}(2007)}]{Dr2007}
\bibinfo{author}{\bibfnamefont{W.}~\bibnamefont{D\"{u}r}} \bibnamefont{and}
  \bibinfo{author}{\bibfnamefont{H.~J.} \bibnamefont{Briegel}},
  \bibinfo{journal}{Reports on Progress in Physics}
  \textbf{\bibinfo{volume}{70}}, \bibinfo{pages}{1381} (\bibinfo{year}{2007}),
  \urlprefix\url{https://doi.org/10.1088/0034-4885/70/8/r03}.

\bibitem[{\citenamefont{Eisert et~al.}(2020)\citenamefont{Eisert, Hangleiter,
  Walk, Roth, Markham, Parekh, Chabaud, and Kashefi}}]{Eisert2020}
\bibinfo{author}{\bibfnamefont{J.}~\bibnamefont{Eisert}},
  \bibinfo{author}{\bibfnamefont{D.}~\bibnamefont{Hangleiter}},
  \bibinfo{author}{\bibfnamefont{N.}~\bibnamefont{Walk}},
  \bibinfo{author}{\bibfnamefont{I.}~\bibnamefont{Roth}},
  \bibinfo{author}{\bibfnamefont{D.}~\bibnamefont{Markham}},
  \bibinfo{author}{\bibfnamefont{R.}~\bibnamefont{Parekh}},
  \bibinfo{author}{\bibfnamefont{U.}~\bibnamefont{Chabaud}}, \bibnamefont{and}
  \bibinfo{author}{\bibfnamefont{E.}~\bibnamefont{Kashefi}},
  \bibinfo{journal}{Nature Reviews Physics} \textbf{\bibinfo{volume}{2}},
  \bibinfo{pages}{382} (\bibinfo{year}{2020}),
  \urlprefix\url{https://doi.org/10.1038/s42254-020-0186-4}.

\bibitem[{\citenamefont{Kliesch and Roth}(2021)}]{Kliesch2021}
\bibinfo{author}{\bibfnamefont{M.}~\bibnamefont{Kliesch}} \bibnamefont{and}
  \bibinfo{author}{\bibfnamefont{I.}~\bibnamefont{Roth}}, \bibinfo{journal}{PRX
  Quantum} \textbf{\bibinfo{volume}{2}}, \bibinfo{pages}{010201}
  (\bibinfo{year}{2021}),
  \urlprefix\url{https://link.aps.org/doi/10.1103/PRXQuantum.2.010201}.

\bibitem[{\citenamefont{G\"uhne et~al.}(2021)\citenamefont{G\"uhne, Mao, and
  Yu}}]{Guhne2021}
\bibinfo{author}{\bibfnamefont{O.}~\bibnamefont{G\"uhne}},
  \bibinfo{author}{\bibfnamefont{Y.}~\bibnamefont{Mao}}, \bibnamefont{and}
  \bibinfo{author}{\bibfnamefont{X.-D.} \bibnamefont{Yu}},
  \bibinfo{journal}{Phys. Rev. Lett.} \textbf{\bibinfo{volume}{126}},
  \bibinfo{pages}{140503} (\bibinfo{year}{2021}),
  \urlprefix\url{https://link.aps.org/doi/10.1103/PhysRevLett.126.140503}.

\bibitem[{\citenamefont{Cruz et~al.}(2019)\citenamefont{Cruz, Fournier,
  Gremion, Jeannerot, Komagata, Tosic, Thiesbrummel, Chan, Macris, Dupertuis
  et~al.}}]{Cruz2019}
\bibinfo{author}{\bibfnamefont{D.}~\bibnamefont{Cruz}},
  \bibinfo{author}{\bibfnamefont{R.}~\bibnamefont{Fournier}},
  \bibinfo{author}{\bibfnamefont{F.}~\bibnamefont{Gremion}},
  \bibinfo{author}{\bibfnamefont{A.}~\bibnamefont{Jeannerot}},
  \bibinfo{author}{\bibfnamefont{K.}~\bibnamefont{Komagata}},
  \bibinfo{author}{\bibfnamefont{T.}~\bibnamefont{Tosic}},
  \bibinfo{author}{\bibfnamefont{J.}~\bibnamefont{Thiesbrummel}},
  \bibinfo{author}{\bibfnamefont{C.~L.} \bibnamefont{Chan}},
  \bibinfo{author}{\bibfnamefont{N.}~\bibnamefont{Macris}},
  \bibinfo{author}{\bibfnamefont{M.-A.} \bibnamefont{Dupertuis}},
  \bibnamefont{et~al.}, \bibinfo{journal}{Advanced Quantum Technologies}
  \textbf{\bibinfo{volume}{2}}, \bibinfo{pages}{1900015}
  (\bibinfo{year}{2019}),
  \urlprefix\url{https://doi.org/10.1002/qute.201900015}.

\bibitem[{\citenamefont{G\"{u}hne and T{\'{o}}th}(2009)}]{Ghne2009}
\bibinfo{author}{\bibfnamefont{O.}~\bibnamefont{G\"{u}hne}} \bibnamefont{and}
  \bibinfo{author}{\bibfnamefont{G.}~\bibnamefont{T{\'{o}}th}},
  \bibinfo{journal}{Physics Reports} \textbf{\bibinfo{volume}{474}},
  \bibinfo{pages}{1} (\bibinfo{year}{2009}),
  \urlprefix\url{https://doi.org/10.1016/j.physrep.2009.02.004}.

\bibitem[{\citenamefont{Horodecki et~al.}(1996)\citenamefont{Horodecki,
  Horodecki, and Horodecki}}]{HORODECKI19961}
\bibinfo{author}{\bibfnamefont{M.}~\bibnamefont{Horodecki}},
  \bibinfo{author}{\bibfnamefont{P.}~\bibnamefont{Horodecki}},
  \bibnamefont{and}
  \bibinfo{author}{\bibfnamefont{R.}~\bibnamefont{Horodecki}},
  \bibinfo{journal}{Physics Letters A} \textbf{\bibinfo{volume}{223}},
  \bibinfo{pages}{1} (\bibinfo{year}{1996}), ISSN \bibinfo{issn}{0375-9601},
  \urlprefix\url{https://www.sciencedirect.com/science/article/pii/S0375960196007062}.

\bibitem[{\citenamefont{Terhal}(2000)}]{TERHAL2000319}
\bibinfo{author}{\bibfnamefont{B.~M.} \bibnamefont{Terhal}},
  \bibinfo{journal}{Physics Letters A} \textbf{\bibinfo{volume}{271}},
  \bibinfo{pages}{319} (\bibinfo{year}{2000}), ISSN \bibinfo{issn}{0375-9601},
  \urlprefix\url{https://www.sciencedirect.com/science/article/pii/S0375960100004011}.

\bibitem[{\citenamefont{Heinosaari and Ziman}(2009)}]{Heinosaari2009}
\bibinfo{author}{\bibfnamefont{T.}~\bibnamefont{Heinosaari}} \bibnamefont{and}
  \bibinfo{author}{\bibfnamefont{M.}~\bibnamefont{Ziman}},
  \emph{\bibinfo{title}{The Mathematical language of Quantum Theory}}
  (\bibinfo{publisher}{Cambridge University Press}, \bibinfo{year}{2009}),
  \urlprefix\url{https://doi.org/10.1017/cbo9781139031103}.

\bibitem[{\citenamefont{Ioannou et~al.}(2004)\citenamefont{Ioannou,
  Travaglione, Cheung, and Ekert}}]{Ioannou2004}
\bibinfo{author}{\bibfnamefont{L.~M.} \bibnamefont{Ioannou}},
  \bibinfo{author}{\bibfnamefont{B.~C.} \bibnamefont{Travaglione}},
  \bibinfo{author}{\bibfnamefont{D.}~\bibnamefont{Cheung}}, \bibnamefont{and}
  \bibinfo{author}{\bibfnamefont{A.~K.} \bibnamefont{Ekert}},
  \bibinfo{journal}{Phys. Rev. A} \textbf{\bibinfo{volume}{70}},
  \bibinfo{pages}{060303} (\bibinfo{year}{2004}),
  \urlprefix\url{https://link.aps.org/doi/10.1103/PhysRevA.70.060303}.

\bibitem[{\citenamefont{Ioannou and Travaglione}(2006)}]{Ioannou2006}
\bibinfo{author}{\bibfnamefont{L.~M.} \bibnamefont{Ioannou}} \bibnamefont{and}
  \bibinfo{author}{\bibfnamefont{B.~C.} \bibnamefont{Travaglione}},
  \bibinfo{journal}{Phys. Rev. A} \textbf{\bibinfo{volume}{73}},
  \bibinfo{pages}{052314} (\bibinfo{year}{2006}),
  \urlprefix\url{https://link.aps.org/doi/10.1103/PhysRevA.73.052314}.

\bibitem[{\citenamefont{Ioannou}(2006)}]{Ioannou2007}
\bibinfo{author}{\bibfnamefont{L.~M.} \bibnamefont{Ioannou}}
  (\bibinfo{year}{2006}),
  \urlprefix\url{https://arxiv.org/abs/quant-ph/0603199}.

\bibitem[{\citenamefont{Bengtsson and
  {\.Z}yczkowski}(2017)}]{bengtsson2017geometry}
\bibinfo{author}{\bibfnamefont{I.}~\bibnamefont{Bengtsson}} \bibnamefont{and}
  \bibinfo{author}{\bibfnamefont{K.}~\bibnamefont{{\.Z}yczkowski}},
  \emph{\bibinfo{title}{Geometry of quantum states: an introduction to quantum
  entanglement}} (\bibinfo{publisher}{Cambridge university press},
  \bibinfo{year}{2017}).

\bibitem[{\citenamefont{D\"ur et~al.}(2000)\citenamefont{D\"ur, Vidal, and
  Cirac}}]{Dur2000}
\bibinfo{author}{\bibfnamefont{W.}~\bibnamefont{D\"ur}},
  \bibinfo{author}{\bibfnamefont{G.}~\bibnamefont{Vidal}}, \bibnamefont{and}
  \bibinfo{author}{\bibfnamefont{J.~I.} \bibnamefont{Cirac}},
  \bibinfo{journal}{Phys. Rev. A} \textbf{\bibinfo{volume}{62}},
  \bibinfo{pages}{062314} (\bibinfo{year}{2000}),
  \urlprefix\url{https://link.aps.org/doi/10.1103/PhysRevA.62.062314}.

\bibitem[{\citenamefont{Verstraete et~al.}(2002)\citenamefont{Verstraete,
  Dehaene, De~Moor, and Verschelde}}]{Verstraete2002}
\bibinfo{author}{\bibfnamefont{F.}~\bibnamefont{Verstraete}},
  \bibinfo{author}{\bibfnamefont{J.}~\bibnamefont{Dehaene}},
  \bibinfo{author}{\bibfnamefont{B.}~\bibnamefont{De~Moor}}, \bibnamefont{and}
  \bibinfo{author}{\bibfnamefont{H.}~\bibnamefont{Verschelde}},
  \bibinfo{journal}{Phys. Rev. A} \textbf{\bibinfo{volume}{65}},
  \bibinfo{pages}{052112} (\bibinfo{year}{2002}),
  \urlprefix\url{https://link.aps.org/doi/10.1103/PhysRevA.65.052112}.

\bibitem[{\citenamefont{Lamata et~al.}(2007)\citenamefont{Lamata, Le\'on,
  Salgado, and Solano}}]{Lamata2007}
\bibinfo{author}{\bibfnamefont{L.}~\bibnamefont{Lamata}},
  \bibinfo{author}{\bibfnamefont{J.}~\bibnamefont{Le\'on}},
  \bibinfo{author}{\bibfnamefont{D.}~\bibnamefont{Salgado}}, \bibnamefont{and}
  \bibinfo{author}{\bibfnamefont{E.}~\bibnamefont{Solano}},
  \bibinfo{journal}{Phys. Rev. A} \textbf{\bibinfo{volume}{75}},
  \bibinfo{pages}{022318} (\bibinfo{year}{2007}),
  \urlprefix\url{https://link.aps.org/doi/10.1103/PhysRevA.75.022318}.

\bibitem[{\citenamefont{Sanz et~al.}(2016)\citenamefont{Sanz, Egusquiza,
  Candia, Saberi, Lamata, and Solano}}]{Sanz2016}
\bibinfo{author}{\bibfnamefont{M.}~\bibnamefont{Sanz}},
  \bibinfo{author}{\bibfnamefont{I.~L.} \bibnamefont{Egusquiza}},
  \bibinfo{author}{\bibfnamefont{R.~D.} \bibnamefont{Candia}},
  \bibinfo{author}{\bibfnamefont{H.}~\bibnamefont{Saberi}},
  \bibinfo{author}{\bibfnamefont{L.}~\bibnamefont{Lamata}}, \bibnamefont{and}
  \bibinfo{author}{\bibfnamefont{E.}~\bibnamefont{Solano}},
  \bibinfo{journal}{Scientific Reports} \textbf{\bibinfo{volume}{6}}
  (\bibinfo{year}{2016}), \urlprefix\url{https://doi.org/10.1038/srep30188}.

\bibitem[{\citenamefont{Hebenstreit et~al.}(2018)\citenamefont{Hebenstreit,
  Gachechiladze, G\"uhne, and Kraus}}]{Gachechiladze2018}
\bibinfo{author}{\bibfnamefont{M.}~\bibnamefont{Hebenstreit}},
  \bibinfo{author}{\bibfnamefont{M.}~\bibnamefont{Gachechiladze}},
  \bibinfo{author}{\bibfnamefont{O.}~\bibnamefont{G\"uhne}}, \bibnamefont{and}
  \bibinfo{author}{\bibfnamefont{B.}~\bibnamefont{Kraus}},
  \bibinfo{journal}{Phys. Rev. A} \textbf{\bibinfo{volume}{97}},
  \bibinfo{pages}{032330} (\bibinfo{year}{2018}),
  \urlprefix\url{https://link.aps.org/doi/10.1103/PhysRevA.97.032330}.

\bibitem[{\citenamefont{Cortes and Vapnik}(1995)}]{cortes1995support}
\bibinfo{author}{\bibfnamefont{C.}~\bibnamefont{Cortes}} \bibnamefont{and}
  \bibinfo{author}{\bibfnamefont{V.}~\bibnamefont{Vapnik}},
  \bibinfo{journal}{Machine learning} \textbf{\bibinfo{volume}{20}},
  \bibinfo{pages}{273} (\bibinfo{year}{1995}).

\bibitem[{\citenamefont{Bishop}(2007)}]{bishop2007}
\bibinfo{author}{\bibfnamefont{C.~M.} \bibnamefont{Bishop}},
  \emph{\bibinfo{title}{Pattern Recognition and Machine Learning (Information
  Science and Statistics)}} (\bibinfo{publisher}{Springer},
  \bibinfo{year}{2007}), \bibinfo{edition}{1st} ed., ISBN
  \bibinfo{isbn}{0387310738}.

\bibitem[{\citenamefont{Carleo and Troyer}(2017)}]{Carleo2017}
\bibinfo{author}{\bibfnamefont{G.}~\bibnamefont{Carleo}} \bibnamefont{and}
  \bibinfo{author}{\bibfnamefont{M.}~\bibnamefont{Troyer}},
  \bibinfo{journal}{Science} \textbf{\bibinfo{volume}{355}},
  \bibinfo{pages}{602} (\bibinfo{year}{2017}),
  \eprint{https://www.science.org/doi/pdf/10.1126/science.aag2302},
  \urlprefix\url{https://www.science.org/doi/abs/10.1126/science.aag2302}.

\bibitem[{\citenamefont{Torlai et~al.}(2018)\citenamefont{Torlai, Mazzola,
  Carrasquilla, Troyer, Melko, and Carleo}}]{Torlai2018}
\bibinfo{author}{\bibfnamefont{G.}~\bibnamefont{Torlai}},
  \bibinfo{author}{\bibfnamefont{G.}~\bibnamefont{Mazzola}},
  \bibinfo{author}{\bibfnamefont{J.}~\bibnamefont{Carrasquilla}},
  \bibinfo{author}{\bibfnamefont{M.}~\bibnamefont{Troyer}},
  \bibinfo{author}{\bibfnamefont{R.}~\bibnamefont{Melko}}, \bibnamefont{and}
  \bibinfo{author}{\bibfnamefont{G.}~\bibnamefont{Carleo}},
  \bibinfo{journal}{Nature Physics} \textbf{\bibinfo{volume}{14}},
  \bibinfo{pages}{447} (\bibinfo{year}{2018}),
  \urlprefix\url{https://doi.org/10.1038/s41567-018-0048-5}.

\bibitem[{\citenamefont{Carleo et~al.}(2019)\citenamefont{Carleo, Cirac,
  Cranmer, Daudet, Schuld, Tishby, Vogt-Maranto, and Zdeborov\'a}}]{RMP2019ML}
\bibinfo{author}{\bibfnamefont{G.}~\bibnamefont{Carleo}},
  \bibinfo{author}{\bibfnamefont{I.}~\bibnamefont{Cirac}},
  \bibinfo{author}{\bibfnamefont{K.}~\bibnamefont{Cranmer}},
  \bibinfo{author}{\bibfnamefont{L.}~\bibnamefont{Daudet}},
  \bibinfo{author}{\bibfnamefont{M.}~\bibnamefont{Schuld}},
  \bibinfo{author}{\bibfnamefont{N.}~\bibnamefont{Tishby}},
  \bibinfo{author}{\bibfnamefont{L.}~\bibnamefont{Vogt-Maranto}},
  \bibnamefont{and}
  \bibinfo{author}{\bibfnamefont{L.}~\bibnamefont{Zdeborov\'a}},
  \bibinfo{journal}{Rev. Mod. Phys.} \textbf{\bibinfo{volume}{91}},
  \bibinfo{pages}{045002} (\bibinfo{year}{2019}),
  \urlprefix\url{https://link.aps.org/doi/10.1103/RevModPhys.91.045002}.

\bibitem[{\citenamefont{Luchnikov et~al.}(2020)\citenamefont{Luchnikov,
  Vintskevich, Grigoriev, and Filippov}}]{LVGF}
\bibinfo{author}{\bibfnamefont{I.~A.} \bibnamefont{Luchnikov}},
  \bibinfo{author}{\bibfnamefont{S.~V.} \bibnamefont{Vintskevich}},
  \bibinfo{author}{\bibfnamefont{D.~A.} \bibnamefont{Grigoriev}},
  \bibnamefont{and} \bibinfo{author}{\bibfnamefont{S.~N.}
  \bibnamefont{Filippov}}, \bibinfo{journal}{Phys. Rev. Lett.}
  \textbf{\bibinfo{volume}{124}}, \bibinfo{pages}{140502}
  (\bibinfo{year}{2020}),
  \urlprefix\url{https://link.aps.org/doi/10.1103/PhysRevLett.124.140502}.

\bibitem[{\citenamefont{Ma and Yung}(2018)}]{Ma2018}
\bibinfo{author}{\bibfnamefont{Y.-C.} \bibnamefont{Ma}} \bibnamefont{and}
  \bibinfo{author}{\bibfnamefont{M.-H.} \bibnamefont{Yung}},
  \bibinfo{journal}{npj Quantum Information} \textbf{\bibinfo{volume}{4}}
  (\bibinfo{year}{2018}),
  \urlprefix\url{https://doi.org/10.1038/s41534-018-0081-3}.

\bibitem[{\citenamefont{Lu et~al.}(2018)\citenamefont{Lu, Huang, Li, Li, Chen,
  Lu, Ji, Shen, Zhou, and Zeng}}]{Lu2018}
\bibinfo{author}{\bibfnamefont{S.}~\bibnamefont{Lu}},
  \bibinfo{author}{\bibfnamefont{S.}~\bibnamefont{Huang}},
  \bibinfo{author}{\bibfnamefont{K.}~\bibnamefont{Li}},
  \bibinfo{author}{\bibfnamefont{J.}~\bibnamefont{Li}},
  \bibinfo{author}{\bibfnamefont{J.}~\bibnamefont{Chen}},
  \bibinfo{author}{\bibfnamefont{D.}~\bibnamefont{Lu}},
  \bibinfo{author}{\bibfnamefont{Z.}~\bibnamefont{Ji}},
  \bibinfo{author}{\bibfnamefont{Y.}~\bibnamefont{Shen}},
  \bibinfo{author}{\bibfnamefont{D.}~\bibnamefont{Zhou}}, \bibnamefont{and}
  \bibinfo{author}{\bibfnamefont{B.}~\bibnamefont{Zeng}},
  \bibinfo{journal}{Phys. Rev. A} \textbf{\bibinfo{volume}{98}},
  \bibinfo{pages}{012315} (\bibinfo{year}{2018}),
  \urlprefix\url{https://link.aps.org/doi/10.1103/PhysRevA.98.012315}.

\bibitem[{\citenamefont{Zhu et~al.}(2021)\citenamefont{Zhu, Wu, Levi, and
  Qian}}]{zhu2021}
\bibinfo{author}{\bibfnamefont{E.~Y.} \bibnamefont{Zhu}},
  \bibinfo{author}{\bibfnamefont{L.~T.} \bibnamefont{Wu}},
  \bibinfo{author}{\bibfnamefont{O.}~\bibnamefont{Levi}}, \bibnamefont{and}
  \bibinfo{author}{\bibfnamefont{L.}~\bibnamefont{Qian}},
  \bibinfo{journal}{arXiv preprint arXiv:2107.02301}  (\bibinfo{year}{2021}).

\bibitem[{\citenamefont{G{\'e}ron}(2019)}]{HandsOn}
\bibinfo{author}{\bibfnamefont{A.}~\bibnamefont{G{\'e}ron}},
  \emph{\bibinfo{title}{Hands-on machine learning with Scikit-Learn, Keras, and
  TensorFlow: Concepts, tools, and techniques to build intelligent systems}}
  (\bibinfo{publisher}{" O'Reilly Media, Inc."}, \bibinfo{year}{2019}).

\bibitem[{\citenamefont{Johansson et~al.}(2012)\citenamefont{Johansson, Nation,
  and Nori}}]{QUTIP2012}
\bibinfo{author}{\bibfnamefont{J.}~\bibnamefont{Johansson}},
  \bibinfo{author}{\bibfnamefont{P.}~\bibnamefont{Nation}}, \bibnamefont{and}
  \bibinfo{author}{\bibfnamefont{F.}~\bibnamefont{Nori}},
  \bibinfo{journal}{Computer Physics Communications}
  \textbf{\bibinfo{volume}{183}}, \bibinfo{pages}{1760} (\bibinfo{year}{2012}),
  ISSN \bibinfo{issn}{0010-4655},
  \urlprefix\url{https://www.sciencedirect.com/science/article/pii/S0010465512000835}.

\bibitem[{\citenamefont{Johansson et~al.}(2013)\citenamefont{Johansson, Nation,
  and Nori}}]{QUTIP2013}
\bibinfo{author}{\bibfnamefont{J.}~\bibnamefont{Johansson}},
  \bibinfo{author}{\bibfnamefont{P.}~\bibnamefont{Nation}}, \bibnamefont{and}
  \bibinfo{author}{\bibfnamefont{F.}~\bibnamefont{Nori}},
  \bibinfo{journal}{Computer Physics Communications}
  \textbf{\bibinfo{volume}{184}}, \bibinfo{pages}{1234} (\bibinfo{year}{2013}),
  ISSN \bibinfo{issn}{0010-4655},
  \urlprefix\url{https://www.sciencedirect.com/science/article/pii/S0010465512003955}.

\bibitem[{\citenamefont{Fasi and Robol}(2020)}]{FasiHaar}
\bibinfo{author}{\bibfnamefont{M.}~\bibnamefont{Fasi}} \bibnamefont{and}
  \bibinfo{author}{\bibfnamefont{L.}~\bibnamefont{Robol}},
  \emph{\bibinfo{title}{Sampling the eigenvalues of random orthogonal and
  unitary matrices}} (\bibinfo{year}{2020}),
  \urlprefix\url{https://arxiv.org/abs/2009.11515}.

\bibitem[{\citenamefont{{TensorFlow Developers}}(2022)}]{TF}
\bibinfo{author}{\bibnamefont{{TensorFlow Developers}}},
  \emph{\bibinfo{title}{Tensorflow}} (\bibinfo{year}{2022}),
  \urlprefix\url{https://zenodo.org/record/4724125}.

\bibitem[{\citenamefont{Huber et~al.}(2021)\citenamefont{Huber, Klep, Magron,
  and Volčič}}]{Huber2021}
\bibinfo{author}{\bibfnamefont{F.}~\bibnamefont{Huber}},
  \bibinfo{author}{\bibfnamefont{I.}~\bibnamefont{Klep}},
  \bibinfo{author}{\bibfnamefont{V.}~\bibnamefont{Magron}}, \bibnamefont{and}
  \bibinfo{author}{\bibfnamefont{J.}~\bibnamefont{Volčič}},
  \emph{\bibinfo{title}{Dimension-free entanglement detection in multipartite
  werner states}} (\bibinfo{year}{2021}),
  \urlprefix\url{https://arxiv.org/abs/2108.08720}.

\bibitem[{\citenamefont{Chru\ifmmode \acute{s}\else
  \'{s}\fi{}ci\ifmmode~\acute{n}\else \'{n}\fi{}ski and
  Kossakowski}(2006)}]{Dariusz2006}
\bibinfo{author}{\bibfnamefont{D.}~\bibnamefont{Chru\ifmmode \acute{s}\else
  \'{s}\fi{}ci\ifmmode~\acute{n}\else \'{n}\fi{}ski}} \bibnamefont{and}
  \bibinfo{author}{\bibfnamefont{A.}~\bibnamefont{Kossakowski}},
  \bibinfo{journal}{Phys. Rev. A} \textbf{\bibinfo{volume}{73}},
  \bibinfo{pages}{062314} (\bibinfo{year}{2006}),
  \urlprefix\url{https://link.aps.org/doi/10.1103/PhysRevA.73.062314}.

\bibitem[{\citenamefont{Abadi et~al.}(2015)\citenamefont{Abadi, Agarwal,
  Barham, Brevdo, Chen, Citro, Corrado, Davis, Dean, Devin
  et~al.}}]{tensorflow2015-whitepaper}
\bibinfo{author}{\bibfnamefont{M.}~\bibnamefont{Abadi}},
  \bibinfo{author}{\bibfnamefont{A.}~\bibnamefont{Agarwal}},
  \bibinfo{author}{\bibfnamefont{P.}~\bibnamefont{Barham}},
  \bibinfo{author}{\bibfnamefont{E.}~\bibnamefont{Brevdo}},
  \bibinfo{author}{\bibfnamefont{Z.}~\bibnamefont{Chen}},
  \bibinfo{author}{\bibfnamefont{C.}~\bibnamefont{Citro}},
  \bibinfo{author}{\bibfnamefont{G.~S.} \bibnamefont{Corrado}},
  \bibinfo{author}{\bibfnamefont{A.}~\bibnamefont{Davis}},
  \bibinfo{author}{\bibfnamefont{J.}~\bibnamefont{Dean}},
  \bibinfo{author}{\bibfnamefont{M.}~\bibnamefont{Devin}},
  \bibnamefont{et~al.}, \emph{\bibinfo{title}{{TensorFlow}: Large-scale machine
  learning on heterogeneous systems}} (\bibinfo{year}{2015}),
  \bibinfo{note}{software available from tensorflow.org},
  \urlprefix\url{https://www.tensorflow.org/}.

\bibitem[{\citenamefont{Pedregosa et~al.}(2011)\citenamefont{Pedregosa,
  Varoquaux, Gramfort, Michel, Thirion, Grisel, Blondel, Prettenhofer, Weiss,
  Dubourg et~al.}}]{pedregosa2011scikit}
\bibinfo{author}{\bibfnamefont{F.}~\bibnamefont{Pedregosa}},
  \bibinfo{author}{\bibfnamefont{G.}~\bibnamefont{Varoquaux}},
  \bibinfo{author}{\bibfnamefont{A.}~\bibnamefont{Gramfort}},
  \bibinfo{author}{\bibfnamefont{V.}~\bibnamefont{Michel}},
  \bibinfo{author}{\bibfnamefont{B.}~\bibnamefont{Thirion}},
  \bibinfo{author}{\bibfnamefont{O.}~\bibnamefont{Grisel}},
  \bibinfo{author}{\bibfnamefont{M.}~\bibnamefont{Blondel}},
  \bibinfo{author}{\bibfnamefont{P.}~\bibnamefont{Prettenhofer}},
  \bibinfo{author}{\bibfnamefont{R.}~\bibnamefont{Weiss}},
  \bibinfo{author}{\bibfnamefont{V.}~\bibnamefont{Dubourg}},
  \bibnamefont{et~al.}, \bibinfo{journal}{Journal of machine learning research}
  \textbf{\bibinfo{volume}{12}}, \bibinfo{pages}{2825} (\bibinfo{year}{2011}).

\bibitem[{\citenamefont{GitHub}(2022)}]{github}
\bibinfo{author}{\bibnamefont{GitHub}}, \emph{\bibinfo{title}{Classification of
  four-qubit entangled states via machine learning: Source code}},
  \bibinfo{howpublished}{\href{https://github.com/StephenVintskevich/qCSC}{https://github.com/StephenVintskevich/qCSC}}
  (\bibinfo{year}{2022}).

\bibitem[{\citenamefont{Stankus et~al.}(2022)\citenamefont{Stankus, Nomerotski,
  Slosar, and Vintskevich}}]{Stankus2022}
\bibinfo{author}{\bibfnamefont{P.}~\bibnamefont{Stankus}},
  \bibinfo{author}{\bibfnamefont{A.}~\bibnamefont{Nomerotski}},
  \bibinfo{author}{\bibfnamefont{A.}~\bibnamefont{Slosar}}, \bibnamefont{and}
  \bibinfo{author}{\bibfnamefont{S.}~\bibnamefont{Vintskevich}},
  \bibinfo{journal}{The Open Journal of Astrophysics}
  \textbf{\bibinfo{volume}{5}} (\bibinfo{year}{2022}),
  \urlprefix\url{https://doi.org/10.21105/astro.2010.09100}.

\bibitem[{\citenamefont{Khabiboulline
  et~al.}(2019{\natexlab{a}})\citenamefont{Khabiboulline, Borregaard, De~Greve,
  and Lukin}}]{LUKIN1}
\bibinfo{author}{\bibfnamefont{E.~T.} \bibnamefont{Khabiboulline}},
  \bibinfo{author}{\bibfnamefont{J.}~\bibnamefont{Borregaard}},
  \bibinfo{author}{\bibfnamefont{K.}~\bibnamefont{De~Greve}}, \bibnamefont{and}
  \bibinfo{author}{\bibfnamefont{M.~D.} \bibnamefont{Lukin}},
  \bibinfo{journal}{Phys. Rev. Lett.} \textbf{\bibinfo{volume}{123}},
  \bibinfo{pages}{070504} (\bibinfo{year}{2019}{\natexlab{a}}),
  \urlprefix\url{https://link.aps.org/doi/10.1103/PhysRevLett.123.070504}.

\bibitem[{\citenamefont{Khabiboulline
  et~al.}(2019{\natexlab{b}})\citenamefont{Khabiboulline, Borregaard, De~Greve,
  and Lukin}}]{LUKIN2}
\bibinfo{author}{\bibfnamefont{E.~T.} \bibnamefont{Khabiboulline}},
  \bibinfo{author}{\bibfnamefont{J.}~\bibnamefont{Borregaard}},
  \bibinfo{author}{\bibfnamefont{K.}~\bibnamefont{De~Greve}}, \bibnamefont{and}
  \bibinfo{author}{\bibfnamefont{M.~D.} \bibnamefont{Lukin}},
  \bibinfo{journal}{Phys. Rev. A} \textbf{\bibinfo{volume}{100}},
  \bibinfo{pages}{022316} (\bibinfo{year}{2019}{\natexlab{b}}),
  \urlprefix\url{https://link.aps.org/doi/10.1103/PhysRevA.100.022316}.

\bibitem[{\citenamefont{Graydon and Appleby}(2016)}]{Graydon2016}
\bibinfo{author}{\bibfnamefont{M.~A.} \bibnamefont{Graydon}} \bibnamefont{and}
  \bibinfo{author}{\bibfnamefont{D.~M.} \bibnamefont{Appleby}},
  \bibinfo{journal}{Journal of Physics A: Mathematical and Theoretical}
  \textbf{\bibinfo{volume}{49}}, \bibinfo{pages}{33LT02}
  (\bibinfo{year}{2016}),
  \urlprefix\url{https://doi.org/10.1088/1751-8113/49/33/33lt02}.

\bibitem[{\citenamefont{Graydon}(2017)}]{Graydon2017conical}
\bibinfo{author}{\bibfnamefont{M.~A.} \bibnamefont{Graydon}},
  \bibinfo{journal}{arXiv preprint arXiv:1703.06800}  (\bibinfo{year}{2017}).

\bibitem[{\citenamefont{Graydon et~al.}(2021)\citenamefont{Graydon,
  Skanes-Norman, and Wallman}}]{Graydon2021}
\bibinfo{author}{\bibfnamefont{M.~A.} \bibnamefont{Graydon}},
  \bibinfo{author}{\bibfnamefont{J.}~\bibnamefont{Skanes-Norman}},
  \bibnamefont{and} \bibinfo{author}{\bibfnamefont{J.~J.}
  \bibnamefont{Wallman}}, \emph{\bibinfo{title}{Clifford groups are not always
  2-designs}} (\bibinfo{year}{2021}),
  \urlprefix\url{https://arxiv.org/abs/2108.04200}.

\bibitem[{\citenamefont{Graydon et~al.}(2022)\citenamefont{Graydon,
  Skanes-Norman, and Wallman}}]{Graydon2022}
\bibinfo{author}{\bibfnamefont{M.~A.} \bibnamefont{Graydon}},
  \bibinfo{author}{\bibfnamefont{J.}~\bibnamefont{Skanes-Norman}},
  \bibnamefont{and} \bibinfo{author}{\bibfnamefont{J.~J.}
  \bibnamefont{Wallman}}, \emph{\bibinfo{title}{Designing stochastic channels}}
  (\bibinfo{year}{2022}), \urlprefix\url{https://arxiv.org/abs/2201.07156}.

\bibitem[{\citenamefont{Erhard et~al.}(2019)\citenamefont{Erhard, Wallman,
  Postler, Meth, Stricker, Martinez, Schindler, Monz, Emerson, and
  Blatt}}]{Erhard2019}
\bibinfo{author}{\bibfnamefont{A.}~\bibnamefont{Erhard}},
  \bibinfo{author}{\bibfnamefont{J.~J.} \bibnamefont{Wallman}},
  \bibinfo{author}{\bibfnamefont{L.}~\bibnamefont{Postler}},
  \bibinfo{author}{\bibfnamefont{M.}~\bibnamefont{Meth}},
  \bibinfo{author}{\bibfnamefont{R.}~\bibnamefont{Stricker}},
  \bibinfo{author}{\bibfnamefont{E.~A.} \bibnamefont{Martinez}},
  \bibinfo{author}{\bibfnamefont{P.}~\bibnamefont{Schindler}},
  \bibinfo{author}{\bibfnamefont{T.}~\bibnamefont{Monz}},
  \bibinfo{author}{\bibfnamefont{J.}~\bibnamefont{Emerson}}, \bibnamefont{and}
  \bibinfo{author}{\bibfnamefont{R.}~\bibnamefont{Blatt}},
  \bibinfo{journal}{Nature Communications} \textbf{\bibinfo{volume}{10}}
  (\bibinfo{year}{2019}),
  \urlprefix\url{https://doi.org/10.1038/s41467-019-13068-7}.

\bibitem[{\citenamefont{Harper et~al.}(2020)\citenamefont{Harper, Flammia, and
  Wallman}}]{Harper2020}
\bibinfo{author}{\bibfnamefont{R.}~\bibnamefont{Harper}},
  \bibinfo{author}{\bibfnamefont{S.~T.} \bibnamefont{Flammia}},
  \bibnamefont{and} \bibinfo{author}{\bibfnamefont{J.~J.}
  \bibnamefont{Wallman}}, \bibinfo{journal}{Nature Physics}
  \textbf{\bibinfo{volume}{16}}, \bibinfo{pages}{1184} (\bibinfo{year}{2020}),
  \urlprefix\url{https://doi.org/10.1038/s41567-020-0992-8}.

\bibitem[{\citenamefont{Winick et~al.}(2021)\citenamefont{Winick, Wallman, and
  Emerson}}]{Joel2021}
\bibinfo{author}{\bibfnamefont{A.}~\bibnamefont{Winick}},
  \bibinfo{author}{\bibfnamefont{J.~J.} \bibnamefont{Wallman}},
  \bibnamefont{and} \bibinfo{author}{\bibfnamefont{J.}~\bibnamefont{Emerson}},
  \bibinfo{journal}{Phys. Rev. Lett.} \textbf{\bibinfo{volume}{126}},
  \bibinfo{pages}{230502} (\bibinfo{year}{2021}),
  \urlprefix\url{https://link.aps.org/doi/10.1103/PhysRevLett.126.230502}.

\bibitem[{\citenamefont{Horodecki et~al.}(2003)\citenamefont{Horodecki,
  Horodecki, Horodecki, Horodecki, Oppenheim, Sen(De), and
  Sen}}]{Horodecki2003b}
\bibinfo{author}{\bibfnamefont{M.}~\bibnamefont{Horodecki}},
  \bibinfo{author}{\bibfnamefont{K.}~\bibnamefont{Horodecki}},
  \bibinfo{author}{\bibfnamefont{P.}~\bibnamefont{Horodecki}},
  \bibinfo{author}{\bibfnamefont{R.}~\bibnamefont{Horodecki}},
  \bibinfo{author}{\bibfnamefont{J.}~\bibnamefont{Oppenheim}},
  \bibinfo{author}{\bibfnamefont{A.}~\bibnamefont{Sen(De)}}, \bibnamefont{and}
  \bibinfo{author}{\bibfnamefont{U.}~\bibnamefont{Sen}},
  \bibinfo{journal}{Phys. Rev. Lett.} \textbf{\bibinfo{volume}{90}},
  \bibinfo{pages}{100402} (\bibinfo{year}{2003}),
  \urlprefix\url{https://link.aps.org/doi/10.1103/PhysRevLett.90.100402}.

\bibitem[{\citenamefont{Gottesman et~al.}(2012)\citenamefont{Gottesman,
  Jennewein, and Croke}}]{Gottesman}
\bibinfo{author}{\bibfnamefont{D.}~\bibnamefont{Gottesman}},
  \bibinfo{author}{\bibfnamefont{T.}~\bibnamefont{Jennewein}},
  \bibnamefont{and} \bibinfo{author}{\bibfnamefont{S.}~\bibnamefont{Croke}},
  \bibinfo{journal}{Phys. Rev. Lett.} \textbf{\bibinfo{volume}{109}},
  \bibinfo{pages}{070503} (\bibinfo{year}{2012}),
  \urlprefix\url{https://link.aps.org/doi/10.1103/PhysRevLett.109.070503}.

\bibitem[{\citenamefont{Nomerotski et~al.}(2020)\citenamefont{Nomerotski,
  Stankus, Slosar, Vintskevich, Andrewski, Carini, Dolzhenko, England,
  Figueroa, Gera et~al.}}]{INSTR}
\bibinfo{author}{\bibfnamefont{A.}~\bibnamefont{Nomerotski}},
  \bibinfo{author}{\bibfnamefont{P.}~\bibnamefont{Stankus}},
  \bibinfo{author}{\bibfnamefont{A.}~\bibnamefont{Slosar}},
  \bibinfo{author}{\bibfnamefont{S.}~\bibnamefont{Vintskevich}},
  \bibinfo{author}{\bibfnamefont{S.}~\bibnamefont{Andrewski}},
  \bibinfo{author}{\bibfnamefont{G.}~\bibnamefont{Carini}},
  \bibinfo{author}{\bibfnamefont{D.}~\bibnamefont{Dolzhenko}},
  \bibinfo{author}{\bibfnamefont{D.}~\bibnamefont{England}},
  \bibinfo{author}{\bibfnamefont{E.}~\bibnamefont{Figueroa}},
  \bibinfo{author}{\bibfnamefont{S.}~\bibnamefont{Gera}}, \bibnamefont{et~al.},
  in \emph{\bibinfo{booktitle}{Optical and Infrared Interferometry and Imaging
  VII}}, edited by \bibinfo{editor}{\bibfnamefont{P.~G.}
  \bibnamefont{Tuthill}},
  \bibinfo{editor}{\bibfnamefont{A.}~\bibnamefont{Mérand}}, \bibnamefont{and}
  \bibinfo{editor}{\bibfnamefont{S.}~\bibnamefont{Sallum}},
  \bibinfo{organization}{International Society for Optics and Photonics}
  (\bibinfo{publisher}{SPIE}, \bibinfo{year}{2020}), vol.
  \bibinfo{volume}{11446}, pp. \bibinfo{pages}{290 -- 306},
  \urlprefix\url{https://doi.org/10.1117/12.2560272}.

\bibitem[{\citenamefont{Zhuang and Zhang}(2019)}]{Zhuang2019}
\bibinfo{author}{\bibfnamefont{Q.}~\bibnamefont{Zhuang}} \bibnamefont{and}
  \bibinfo{author}{\bibfnamefont{Z.}~\bibnamefont{Zhang}},
  \bibinfo{journal}{Phys. Rev. X} \textbf{\bibinfo{volume}{9}},
  \bibinfo{pages}{041023} (\bibinfo{year}{2019}),
  \urlprefix\url{https://link.aps.org/doi/10.1103/PhysRevX.9.041023}.

\bibitem[{\citenamefont{Ghosh et~al.}(2019)\citenamefont{Ghosh, Opala,
  Matuszewski, Paterek, and Liew}}]{Ghosh2019}
\bibinfo{author}{\bibfnamefont{S.}~\bibnamefont{Ghosh}},
  \bibinfo{author}{\bibfnamefont{A.}~\bibnamefont{Opala}},
  \bibinfo{author}{\bibfnamefont{M.}~\bibnamefont{Matuszewski}},
  \bibinfo{author}{\bibfnamefont{T.}~\bibnamefont{Paterek}}, \bibnamefont{and}
  \bibinfo{author}{\bibfnamefont{T.~C.~H.} \bibnamefont{Liew}},
  \bibinfo{journal}{npj Quantum Information} \textbf{\bibinfo{volume}{5}}
  (\bibinfo{year}{2019}),
  \urlprefix\url{https://doi.org/10.1038/s41534-019-0149-8}.

\bibitem[{\citenamefont{Vintskevich and
  Grigoriev}(2022)}]{vintskevich2022computing}
\bibinfo{author}{\bibfnamefont{S.}~\bibnamefont{Vintskevich}} \bibnamefont{and}
  \bibinfo{author}{\bibfnamefont{D.}~\bibnamefont{Grigoriev}},
  \bibinfo{journal}{arXiv preprint arXiv:2201.07969}  (\bibinfo{year}{2022}).

\end{thebibliography}

\end{document}